\documentclass[journal,twoside,10pt]{IEEEtran}

\usepackage{cite}
\usepackage[pdftex]{graphicx}
\graphicspath{{Fig/}}
\usepackage[cmex10]{amsmath}
\usepackage{amsthm}
\usepackage{amssymb}
\usepackage{multirow}
\usepackage{booktabs}

\usepackage{algorithm}
\usepackage{algpseudocode}

\usepackage{array}
\usepackage{color}
\usepackage{epstopdf}
\usepackage{amsfonts}
\usepackage[acronym,shortcuts]{glossaries}
\usepackage[normalem]{ulem}

\usepackage{pgfplots}
\pgfplotsset{compat=1.18}
\usepackage{tikz}
\usetikzlibrary{shapes}
\usetikzlibrary{spy}
\usetikzlibrary{circuits}
\usetikzlibrary{arrows}
\usetikzlibrary{patterns, decorations.pathreplacing, arrows.meta}
\usepackage{svg}
\usepackage{cancel}
\usepackage{balance}
\usepackage{setspace}

\usepackage{upgreek}
\usepgfplotslibrary{groupplots,fillbetween}
\pgfplotsset{grid style={densely dotted,gray}}
\definecolor{mypink}{RGB}{255, 20, 147}

\newcommand{\vect}[1]{\boldsymbol{\mathrm{#1}}}
\newcommand{\mat}[1]{\boldsymbol{\mathrm{#1}}}

\newcommand{\hr}[1]{#1^{{\scriptscriptstyle\mathrm{H}}}}
\newcommand{\Ma}{M_{\mathrm{a}}}
\newcommand{\Mamin}{M_{\mathrm{a,min}}}
\newcommand{\Na}{N_{\mathrm{a}}}
\newcommand{\Namin}{N_{\mathrm{a,min}}}
\newcommand{\Pa}{P_{\mathrm{a}}}
\newcommand{\Pamin}{P_{\mathrm{a,min}}}
\newcommand{\Pcons}{P_{\mathrm{cons}}}
\newcommand{\Psleep}{P_{\mathrm{sleep}}}
\newcommand{\Pmax}{P_{\mathrm{max}}}
\newcommand*{\inC}[1]{\in\mathbb{C}^{#1}}

\newcommand{\tr}{\mathrm{tr}}

\makeglossaries
\newacronym{ict}{ICT}{information and communications technology}
\newacronym[plural=BSs,firstplural=base stations (BSs)]{bs}{BS}{base station}
\newacronym[plural=PAs,firstplural=power amplifiers (PAs)]{pa}{PA}{power amplifier}
\newacronym{sdr}{SDR}{signal-to-distortion ratio}
\newacronym{snr}{SNR}{signal-to-noise ratio}
\newacronym{sinr}{SINR}{signal-to-interference-plus-noise ratio}
\newacronym{sndr}{SNDR}{signal-to-noise-and-distortion ratio}
\newacronym{snidr}{SNIDR}{signal-to-noise-and-interference-and-distortion ratio}
\newacronym{los}{LOS}{line-of-sight}
\newacronym{z3ro}{Z3RO}{zero third-order distortion}
\newacronym{mimo}{MIMO}{multiple-input multiple-output}
\newacronym{siso}{SISO}{single-input single-output}
\newacronym{mrt}{MRT}{maximum ratio transmission}
\newacronym{dpd}{DPD}{digital pre-distortion}
\newacronym{ula}{ULA}{uniform linear array}
\newacronym{se}{SE}{spectral efficiency}
\newacronym{ee}{EE}{energy efficiency}
\newacronym{tdma}{TDMA}{time-division multiple access}
\newacronym{ran}{RAN}{radio access network}
\newacronym{lte}{LTE}{Long-Term Evolution}
\newacronym{nr}{NR}{New Radio}
\newacronym{sdma}{SDMA}{space-division multiple access}
\newacronym{csi}{CSI}{channel state information}

\newacronym{psd}{PSD}{power spectral density}
\newacronym{aclr}{ACLR}{adjacent channel leakage ratio}
\newacronym{cpu}{CPU}{central processing unit}
\newacronym{iid}{i.i.d.}{independent and identically distributed}
\newacronym{ue}{UE}{user equipment}
\newacronym{nlos}{NLoS}{non-line-of-sight}
\newacronym{hpbm}{HPBM}{half power beam width}
\newacronym{rts}{RTS}{ray tracing simulator}
\newacronym{ag}{AG}{array gain}
\newacronym{arp}{ARP}{Antenna Reference Point}
\newacronym{ecdf}{eCDF}{Empirical Cumulative Distribution Function}
\newacronym[plural=CDFs,firstplural=cumulative distribution functions (CDFs)]{cdf}{CDF}{cumulative distribution function}
\newacronym{evm}{EVM}{Error Vector Magnitude}
\newacronym{fft}{FFT}{fast Fourier transform}
\newacronym{ifft}{IFFT}{inverse fast Fourier transform}
\newacronym{ib}{IB}{in-band}
\newacronym{im}{IM}{intermodulation}
\newacronym{mf}{MF}{matched filtering}
\newacronym{mr}{MR}{maximum ratio}
\newacronym{ofdm}{OFDM}{orthogonal frequency-division multiplexing}
\newacronym{oob}{OOB}{out-of-band}
\newacronym{papr}{PAPR}{peak-to-average power ratio}
\newacronym{rx}{RX}{Receiver}
\newacronym{trp}{TRP}{Transmission Reception Point}
\newacronym{tx}{TX}{transmitter}
\newacronym{zf}{ZF}{zero forcing}
\newacronym{dl}{DL}{downlink}
\newacronym{rzf}{RZF}{regularized zero forcing}
\newacronym{slc}{SLC}{spatial leakage suppression}
\newacronym{kpi}{KPI}{key performance indicator}
\newacronym{awgn}{AWGN}{additive white gaussian noise}
\newacronym{qam}{QAM}{quadrature amplitude modulation}
\newacronym{amam}{AM/AM}{amplitude modulation to amplitude modulation}
\newacronym{ampm}{AM/PM}{amplitude modulation to phase modulation}
\newacronym{zmcscg}{ZMCSCG}{zero mean circularly symmetric complex Gaussian}

\newacronym[plural=FOCs,firstplural=first-order conditions (FOCs)]{foc}{FOC}{first-order condition}
\newacronym{3gpp}{3GPP}{3rd Generation Partnership Project}
\newacronym{5G}{5G}{fifth-generation}
\newacronym{qos}{QoS}{quality of service}
\newacronym{afe}{AFE}{analog front-end}
\newacronym{dbb}{DBB}{digital baseband}
\newacronym{ru}{RU}{radio unit}
\newacronym{rru}{RRU}{remote radio unit}
\newacronym{aau}{AAU}{active antenna unit}
\newacronym{fdd}{FDD}{frequency-division duplexing}
\newacronym{tdd}{TDD}{time-division duplexing}
\newacronym[plural=CQIs]{cqi}{CQI}{channel quality indicator}
\newacronym{dtx}{DTX}{discontinuous transmission}
\newacronym{mudtx}{\textmu DTX}{micro-discontinuous transmission}
\newacronym{1D}{1D}{one-dimensional}
\newacronym{2D}{2D}{two-dimensional}

\usepackage{mathtools}
\DeclarePairedDelimiter\ceil{\lceil}{\rceil}
\DeclarePairedDelimiter\floor{\lfloor}{\rfloor}
\newcommand{\ceilfloor}[1]{\left\lfloor{#1}\right\rceil}
\newcommand\round[1]{\left[{#1}\right]}

\theoremstyle{definition}

\ifCLASSOPTIONcompsoc
  \usepackage[caption=false,font=normalsize,labelfont=sf,textfont=sf]{subfig}
\else
  \usepackage[caption=false,font=footnotesize]{subfig}
\fi
\usepackage{dblfloatfix}

\usepackage[hidelinks]{hyperref}
\newcommand{\ie}{\textit{i.e.}}
\newcommand{\eg}{\textit{e.g.}}

\DeclareMathOperator*{\argmin}{arg\,min}
\usepackage{enumitem}

\begin{document}

\title{\fontsize{21}{\baselineskip}\selectfont \vspace{-.5em}On Optimizing Time-, Space- and Power-Domain Energy-Saving Techniques for Sub-6 GHz Base Stations}
 
\author{
	Emanuele Peschiera,~\IEEEmembership{Graduate Student Member,~IEEE},
	Youssef Agram,
	Fran\c{c}ois Quitin,~\IEEEmembership{Member,~IEEE},\\
	Liesbet Van der Perre,~\IEEEmembership{Senior Member,~IEEE},
	Fran\c{c}ois Rottenberg,~\IEEEmembership{Member,~IEEE}
\vspace{-.5cm}
	\thanks{E. Peschiera, L. Van der Perre and F. Rottenberg are with ESAT-DRAMCO, Campus Ghent, KU Leuven, 9000 Ghent, Belgium (email: emanuele.peschiera@kuleuven.be).
	
	Y. Agram and F. Quitin are with Brussels School of Engineering, ULB, 1000 Brussels, Belgium.}}

\markboth{}{}
%



\maketitle

\begin{abstract}
What is the optimal \gls{bs} resource allocation strategy given a measurement-based power consumption model and a fixed target user rate? Rush-to-sleep in time, rush-to-mute in space, awake-but-whisper in power, or a combination of them? We propose in this paper an efficient solution to the problem of finding the optimal number of active time slots, active antennas, and transmit power at active antennas in a \gls{mimo} \gls{ofdm} system under per-user rate and per-antenna transmit power constraints. The use of a parametric power consumption model validated on operator measurements of 4G and 5G \glspl{bs} enhances the interpretation of the results. We discuss the optimal energy-saving strategy at different network loads for three \gls{bs} configurations. Using as few \gls{bs} antennas as possible is close to optimal in \glspl{bs} not implementing time-domain power savings such as \gls{mudtx}. Energy-saving schemes that jointly operate in the three domains are instead optimal when the \gls{bs} hardware can enter time-domain power-saving modes, with a tendency for rush-to-mute in massive \gls{mimo} and for rush-to-sleep in \gls{bs} with fewer antennas. Median energy savings up to $\mathbf{30\boldsymbol{\%}}$ are achieved at low network loads.
\end{abstract}
\glsresetall


%
\IEEEpeerreviewmaketitle


\vspace{-0.4cm}
\section{Introduction}\label{section:introduction}

\subsection{Motivation}
The reduction of energy consumption in wireless networks is imperative to reach the environmental sustainability objectives and control the expenses of mobile operators~\cite{GSMA_5G_EE,JRC135926}, 
which foresee a mobile data traffic exceeding 400 EB/month in 2029~\cite{Ericsson_mr}.
Most of the energy consumption in a wireless network comes from the \gls{ran} and in particular from the \glspl{bs}~\cite{Gruber_2009, GSMA_5G_EE, nokia_ee_whitepaper}.
Measurements indicate that the power consumption of current wireless networks decreases only slightly at low traffic loads~\cite{golard_evaluation_2022}, 
hence network energy-saving techniques are becoming essential in wireless systems~\cite{3gpp.38.864}.
Energy-saving techniques implement load-aware resource adaptations and \gls{bs} sleep modes to break down consumption at low and medium loads~\cite{islam_enabling_2023},
where definitions of load include the fraction of active \gls{ofdm} time slots, the utilized fraction of the network peak capacity, and the ratio of transmit power to maximum transmit power.

\subsection{State of the Art}
Energy-saving techniques can generally operate in four domains, \ie, time, frequency, space, and power~\cite{islam_enabling_2023,Laselva_2024}. 
The adaptive shutdown of transmit symbols, (sub)carriers, antennas, as well as the adaptation of the transmit power at the \gls{bs}, are among the main drivers to achieve power consumption savings.
The use of accurate power consumption models becomes a crucial aspect~\cite{Lopez2022}. A review of the published models up to 2023 is given in~\cite{Busch_2024}. 
These are classified in (i) linear models if they express the power consumption as linearly dependent on the \gls{bs} output power,
(ii) subcomponent models if they estimate the consumption of individual processing blocks, and (iii) massive \gls{mimo} models when they target 5G \glspl{bs} with 32 or more antennas.
A recent work has proposed a parametric power consumption model able to predict the power consumption of sub-6 GHz 4G and 5G \glspl{bs} in Belgium~\cite{Golard_2024}.
By building upon measurements of deployed network and datasheets of manufacturers, the authors model the consumption of 
\glspl{pa}, \gls{afe}, \gls{dbb}, and power supply and cooling systems. Another recent work introduced the concept of waste factor, which quantifies the
proportion of wasted power versus power effectively used for any intended function, evaluating therefore the total \gls{bs} power consumption~\cite{rappaport_2024}.

As anticipated, energy savings can be achieved by optimizing the resource usage in different domains. 
In the time domain, \gls{mudtx} has been used to switch off \glspl{pa} on a symbol scale~\cite{Frenger_2011,Cheng_2014}.
The 5G networks have been designed to be lean, requiring less frequent control signaling and therefore enabling longer sleep 
durations~\cite{dahlman20205g}. The impact of using sleep modes during both data transmission and control signaling phases has been evaluated at standard 
level~\cite{3gpp.38.864} and via system-level simulations~\cite{Salem_2017,frenger_more_2019}. The tradeoff between energy saving and latency has been 
discussed in~\cite{Renga_2023}. In~\cite{Liu_2023}, the authors introduce a two-dimensional modeling of the \gls{ee} that incorporates the time dimension.
However, fundamental works in this area have often considered that the transmission takes place at a constant rate per time 
slot~\cite{budzisz_dynamic_2014,Zhang_2017}, preventing the assessment of the optimal transmit powers and rates at different time slots.
The work in~\cite{rottenberg2023informationtheoretic} fills this theoretical gap in the case of a \gls{siso} system by considering an average rate 
constraint over multiple time slots. It shows that transmission at constant power among a number of active time slots is optimal and that a 
rush-to-sleep approach, using as few time slots as possible at maximum transmit power, is optimal at low \gls{snr}.

In the space domain, many works have focused on load-aware activation of antennas in massive \gls{mimo} \glspl{bs}~\cite{Lopez2022}. 
In general, the need for these energy-saving schemes becomes evident when optimizing total consumed power instead of transmit power only. 
The authors of~\cite{Bjornson_2015} find the optimal number of active antennas and users that maximize the \gls{ee} by building on a 
detailed circuit consumption model. A similar analysis is done in~\cite{hossain_energy_2018} by considering a multi-cell network and traffic load variations. 
The work~\cite{Senel_2019} addresses the problem of minimizing power consumption under per-user \gls{sinr} constraints.  
The precoder that minimizes \gls{bs} consumption under \gls{sinr} constraints for a multicarrier massive \gls{mimo} system is derived in~\cite{Peschiera2023}. 
It is shown that conventional \gls{zf} among a subset of antennas is optimal. The work~\cite{Rajapaksha_2023} specifically targets energy consumption reduction 
by maximizing the number of muted antennas subject to \gls{qos} constraints and solves the problem using neural networks.
Space-domain resource adaptation becomes even more important with the increasing interest in cell-free \gls{mimo}, 
where it is essential to keep the total consumption under control by turning on/off access points and their antennas~\cite{VanChien_2020,Jayaweera_2024,Yan_2025}.

Power-domain energy-saving techniques can instead optimize either the total \gls{bs} transmit power as is the case in classic precoding designs~\cite{Shenouda_2007},
or the individual transmit powers at the \gls{bs} antennas in order to improve the \glspl{pa} efficiency~\cite{Joung_2014}.
Constant-envelope precoding has been proposed in~\cite{Mohammed_2013}. The authors in~\cite{Cheng_2019} find the optimal point-to-point \gls{mimo} precoder that 
maximizes the rate under a \gls{pa} consumption constraint. More recently, flat-\gls{zf} precoding has been proposed with the same objective of 
reducing \gls{pa} consumption\cite{Sohrabi_2025}.

Recent works have also looked at multi-domain resource allocation. The joint optimization of spectral and spatial resources in a multi-cell 
network under \gls{zf} precoding is addressed in~\cite{Marwaha_2023}, while optimal precoders and scheduling of resources are found in~\cite{Wei_2023} via convex solvers.
In~\cite{Zhang_2025}, the authors optimize the sleep mode configuration, number of active antennas, and user-\gls{bs} associations in a multi-cell network using reinforcement learning.

\begin{table}[t!]\centering
\renewcommand{\arraystretch}{1.25}
\caption{Main quantities and relative definitions.}\label{tab:main_variables}
\vspace{-0.5em}
\resizebox{.5\textwidth}{!}{
\setlength{\tabcolsep}{3pt}
\begin{tabular}{@{}ll@{}}
\toprule
\textbf{Variable} & \textbf{Description}\\
\midrule
$M$, $N$, $K$, $Q$ & Number of antennas, time slots, users, and subcarriers\\
$\Ma$, $\Na$ & Number of active antennas and active time slots\\
$M_\mathrm{a,min}$, $N_\mathrm{a,min}$ & Minimal number of active antennas and active time slots\\
$\Pa$, $P_\mathrm{a,min}$ & Transmit and minimal transmit power at active antennas\\
$\Pmax$, $\alpha$ & Maximum PA output power, PA consumption exponent\\
$\eta_\mathrm{\scriptscriptstyle{PA}}^\mathrm{max}$, $\xi$ & PA maximum efficiency and weight of static consumption\\
$p_k$, $\rho_k$ & Power allocation and single-antenna equiv.~SNR at user $k$\\
$f_\mathrm{c}$, $B$ & Carrier frequency and bandwidth\\
$R_{k,0}$ & Baseline target rate of user $k$\\
$R_k=\kappa R_{k,0}$, $\kappa$ & Targer rate of user $k$, rate scaling\\
$\sigma_k^2$, $\beta_k$ & Noise power and large-scale fading coefficient at user $k$\\
$\tau_\mathrm{\scriptscriptstyle{DL}}$, $\tau_\mathrm{\scriptscriptstyle{UL}}$ & Time ratios of downlink and uplink to total frame\\
$\tau_\mathrm{sig}$ & Time ratio of reference signaling to downlink\\
$\delta_\mathrm{\scriptscriptstyle{PA}}^\mathrm{dtx}$, $\delta_\mathrm{\scriptscriptstyle{PA}}^\mathrm{idle}$, $\delta_\mathrm{\scriptscriptstyle{PA}}^\mathrm{sleep}$ & 
Reduction factors for PA \textmu DTX, idle, and sleep modes\\
$\delta_\mathrm{\scriptscriptstyle{TRX}}^\mathrm{idle}$, $\delta_\mathrm{\scriptscriptstyle{TRX}}^\mathrm{sleep}$ & Reduction factors for AFE idle and sleep modes\\
$\delta_\mathrm{phy}^\mathrm{idle}$ & Reduction factor for DBB idle mode\\
$\bar{P}_\mathrm{\scriptscriptstyle{PA}\scriptstyle{,1}}$, $\bar{P}_\mathrm{\scriptscriptstyle{AFE}\scriptstyle{,1}}$ & PA and AFE contributions to $P_1$\\
$\bar{P}_\mathrm{\scriptscriptstyle{AFE}\scriptstyle{,sleep}}$, $\bar{P}_\mathrm{\scriptscriptstyle{DBB}}$ & AFE contribution to $\Psleep$, DBB consumption\\
$\eta_\mathrm{s/c}^\mathrm{\scriptscriptstyle{PA}}$, $\eta_\mathrm{s/c}^\mathrm{\scriptscriptstyle{AFE}}$, $\eta_\mathrm{s/c}^\mathrm{\scriptscriptstyle{DBB}}$ &
Supply and cooling efficiencies of PA, AFE, and DBB\\
$\gamma$, $P_0$, $P_1$, $\Psleep$ & Parameters of power consumption model\\
\bottomrule
\vspace{-3em}
\end{tabular}
}
\end{table}

\subsection{Contributions}
In this paper, we answer the question of finding the optimal combination of power-, space- and time-domain energy-saving techniques 
for different \gls{bs} configurations at varying network loads. 
Three specific allocations are identified and used as benchmark: a rush-to-sleep
that minimizes the number of active time slots, a rush-to-mute that minimizes the number of active antennas, and an awake-but-whisper that
minimizes the transmit power at active antennas. 
The relative performance of these three allocations is not addressed by previous works, and therefore we do not consider frequency-domain 
energy savings to limit the scope.
We employ a power consumption model applicable to deployed sub-6 GHz \glspl{bs}
and we further use on-site measurements to initialize our numerical simulations~\cite{Golard_2024,Agram_2024}.
Compared to the conference version~\cite{spawc_2025}, we provide here the full expressions of the power consumption
model parameters, the proofs of convexity of the optimization problem, an approximation of the optimal solution and the analysis of 4G and 5G BSs with a lower number of antennas.
Our specific contributions are the following:
\begin{itemize}[leftmargin=*]
    \item We first show that the initial problem can be simplified, casted into a two-dimensional convex differentiable problem (with a sufficient condition on the
    domain convexity) and efficiently solved. We propose an approximation of the final discrete solution and provide the error decay.
    \item We discuss the optimal energy-saving strategy for the 4G and 5G \glspl{bs} under study in~\cite{Golard_2024}, not implementing time-domain power savings
    (PA \gls{mudtx} and \gls{afe} idle modes). A rush-to-mute approach provides quasi-optimal performance at any network load for massive \gls{mimo}
    \glspl{bs} and at low and medium network loads for \glspl{bs} with fewer antennas.
    \item We consider the case when the analyzed 4G and 5G \glspl{bs} utilize time-domain power savings in the form of \gls{pa} \gls{mudtx} and \gls{afe} idle modes.
    It turns out that a combination of the three energy-saving techniques is the optimal one. Massive \gls{mimo} \glspl{bs} still tends more to a rush-to-mute at any network load,
    while the \glspl{bs} with fewer antennas steer towards a rush-to-sleep at low and medium loads.
\end{itemize}

The remainder of the paper is organized as follows. Section~\ref{section:transmission_model} introduces the transmission model in frequency, time, and space. 
Section~\ref{section:power_model} explains the \gls{bs} power consumption model. The solution to the considered optimization problem is discussed in Section~\ref{section:optimal_allocation}.
Numerical experiments are presented in Section~\ref{section:evaluation}. Section~\ref{section:conclusion} concludes the paper.
Table~\ref{tab:main_variables} summarizes the main quantities used throughout the paper.

{\textbf{Notations}}: The symbols $\mathbb{E}\{\cdot\}$ and $\tr[\cdot]$ denote the expectation of a random variable and the trace of a matrix.
The operators $\ceil{\cdot}$, $\floor{\cdot}$, and $\round{\cdot}$ are the ceil, floor, and round, respectively. 
The operator $\ceilfloor{x}$, which we refer to as the ceil-floor, selects among the upper and lower bounding integers of $x$ the one that optimizes the cost function. 
In two dimensions, $\ceilfloor{x,y}$ selects among the four different combinations of $(x,y)$ that use ceil or floor for both variables, the one that optimizes the cost function. 
We use the notation $f(x)=\mathcal{O}(g(x))$, as $x\to a$, if there exist positive numbers $\mu$ and $\lambda$ such that $|f(x)|\leq \lambda g(x)$ when $0<|x-a|<\mu$.

\section{Transmission Model}\label{section:transmission_model}

\begin{figure}[t!]\centering
	\resizebox{.46\textwidth}{!}{
	\includegraphics{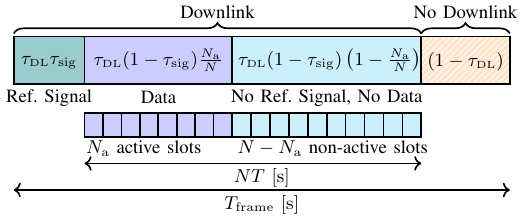}
	}
	\vspace{-0.3cm}
	\caption{Frame structure in time considered in this work, where the quantities inside each block are time ratios relative to the frame duration $T_\mathrm{frame}$.}
	\label{fig:frame}
\end{figure}

We consider a \gls{bs} with $M$ antennas and $K$ single-antenna users communicating in downlink using \gls{ofdm} with $Q$ active subcarriers carrying data symbols. 
The cyclic prefix is assumed to be larger than the channel excess delay. The \gls{ofdm} symbol duration, including the cyclic prefix, is denoted by $T$.
The users are multiplexed using \gls{sdma} implying that they are being served at the same time and frequency, and we consider $M\geq K$.
The transmission time is divided into frames lasting $T_\mathrm{frame}$ seconds. In every frame there are $N$ \gls{ofdm} symbols available to carry data symbols.
Out of the $N$ \gls{ofdm} symbols, only $\Na$ are active, \ie, used to communicate.\footnote{We interchangeably use the terms \gls{ofdm} symbol and time slot.}
A frame is divided in four phases:
\begin{itemize}[leftmargin=*]
\item Downlink and reference signal transmission, lasting $\tau_\mathrm{\scriptscriptstyle{DL}}\tau_\mathrm{sig}T_\mathrm{frame}$ seconds, where 
$\tau_{\scriptscriptstyle{\mathrm{DL}}}\in(0,1]$ is the time ratio of downlink transmission to the frame duration and $\tau_{\mathrm{sig}}\in(0,1)$ is the time ratio of 
reference signal transmission to downlink transmission.
\item Downlink and data transmission, with a duration of $\tau_\mathrm{\scriptscriptstyle{DL}}\left(1-\tau_\mathrm{sig}\right)\frac{\Na}{N}T_\mathrm{frame}$ seconds.
\item Downlink and no reference signal and no data transmission, lasting 
$\tau_\mathrm{\scriptscriptstyle{DL}}\left(1-\tau_\mathrm{sig}\right)\left(1-\frac{\Na}{N}\right)T_\mathrm{frame}$ seconds. 
\item No downlink that spans $\left(1-\tau_\mathrm{\scriptscriptstyle{DL}}\right)T_\mathrm{frame}$ seconds. This is either uplink transmission in \gls{tdd} or 
has zero duration in \gls{fdd} as uplink takes place on a different frequency.
\end{itemize}
The corresponding frame structure is given in Fig.~\ref{fig:frame}. We illustrate the total duration of the four phases, but the reference signals as well as the active and non-active
\gls{ofdm} data symbols can be distributed across the frame. Without loss of generality, we consider one such frame in the following.

Throughout this work, we consider energy-saving techniques where only a subset of $\Ma$ antennas out of $M$ are active, with $K\leq\Ma\leq M$. 
The transmitted symbols at each subcarrier and \gls{ofdm} symbol are considered zero mean and uncorrelated among users. 
There is one symbol per user at a given subcarrier and active \gls{ofdm} symbol, with variance $p_k$ that is the power allocation of user $k$.
We denote the transmitted symbol vector as $\vect{s}\inC{K\times 1}$. 
The corresponding channel matrix between the active antennas and the users is denoted by $\mat{H}\inC{K\times \Ma}$. 
We consider the use of a \gls{zf} precoder so that the precoded symbols are $\hr{\mat{H}}(\mat{H}\hr{\mat{H}})^{-1}\vect{s}$. 
The large-scale fading coefficient of user $k$ is denoted by $\beta_k$. 
While the \gls{iid} Rayleigh fading is commonly used in the massive \gls{mimo} literature~\cite{marzetta_fundamentals_2016}, 
several experimental studies have shown that it is often not verified in practice~\cite{gao_massive_2015,willhammar_channel_2020}. 
Still, this assumption is made here as being useful for the purpose of finding tractable resource allocation strategies. 
Considering a unitary \gls{ifft}, the average power of the precoded symbols in the frequency domain is the same as in the time domain (after the \gls{ifft}). 
Under \gls{iid} Rayleigh fading, the average total transmitted power is then given by
\begin{align}
	P_{\scriptscriptstyle{\mathrm{T}}} &= \mathbb{E}\left\{\tr\!\left[\hr{\mat{H}}(\mat{H}\hr{\mat{H}})^{-1}\vect{s}\hr{\vect{s}}(\mat{H}\hr{\mat{H}})^{-1}\mat{H}\right]\right\}\nonumber \\
	&\stackrel{(*)}{=} \tr\!\left[\mathbb{E}\left\{(\mat{H}\hr{\mat{H}})^{-1}\right\}\mathbb{E}\left\{\vect{s}\hr{\vect{s}}\right\}\right]
	\stackrel{(**)}{=} \frac{1}{\Ma-K}\sum_{k=1}^{K}\frac{p_k}{\beta_k} \label{eq:array_gain}
\end{align}
where $(*)$ follows from the trace cyclic property and the fact that the symbols and channel coefficients are uncorrelated, while $(**)$ uses the expression 
of the mean of a complex inverse-Wishart distribution~\cite{maiwald_moments_1997}. 

An array gain of a factor $\Ma-K$ is present in~(\ref{eq:array_gain}). To minimize $P_{\scriptscriptstyle{\mathrm{T}}}$, the maximal number of antennas should be active, \ie, $\Ma=M$. 
In terms of consumed power, denoted by $\Pcons$, this is often not optimal as $\Pcons$ is not linearly proportional to $P_{\scriptscriptstyle{\mathrm{T}}}$ because of the \gls{pa} non-constant efficiency 
and the load-independent consumption of activating each antenna. This is further detailed in Section~\ref{section:power_model}. 
Given that all antennas are statistically identically distributed, the average transmit power is the same at each antenna and equal to 
\begin{align}\label{eq:general_Pa_expression}
	\Pa = \frac{P_{\scriptscriptstyle{\mathrm{T}}}}{\Ma} = \frac{1}{\Ma(\Ma-K)}\sum_{k=1}^{K}\frac{p_k}{\beta_k}.
\end{align}
The selection of the $\Ma$ active antennas can be arbitrary given their identical distribution. In the following, we consider a maximal per-antenna transmit power constraint $\Pa\leq \Pmax$.
After going through the channel and \gls{ofdm} demodulation at the receiver, the transmitted symbols are recovered with the addition of zero-mean circularly-symmetric complex Gaussian noise. 
The noise variance at user $k$ is denoted by $\sigma_k^2$. Then, the total bit rate delivered to user $k$
averaged over the frame duration is given by
\begin{align}\label{eq:rate}
	R_{k,\mathrm{deliv}} &= \frac{Q}{T}\tau_\mathrm{\scriptscriptstyle{DL}}\left(1-\tau_\mathrm{sig}\right)\frac{\Na}{N}\log_2\left(1+\frac{p_{k}}{\sigma_k^2}\right)
\end{align}
expressed in bits per second. 
To simplify notations, we define $R_k=R_{k,\mathrm{deliv}}/\big(\frac{Q}{T}\tau_\mathrm{\scriptscriptstyle{DL}}(1-\tau_\mathrm{sig})\big)$ as
the normalized per-subcarrier average rate expressed in bits per \gls{ofdm} symbol and per subcarrier.
In the following, we consider that each user has a target average rate $R_k$. 
Given that we analyse one frame transmission, this rate constraint can also be seen as a latency constraint.
If we consider that we first send all reference signals and then all the active \gls{ofdm} symbols, we are constraining the 
delivery of $NTR_{k,\mathrm{deliv}}$ bits to the $k$-th user in 
$\tau_\mathrm{\scriptscriptstyle{DL}}\tau_\mathrm{sig}T_\mathrm{frame}+NT$ seconds at most.

\begin{table*}[t!]\centering\label{tab:parameters_Pcons}
\caption{Parameters of power consumption model~(\ref{eq:P_cons_general}) for different \gls{bs} configurations.}
\vspace{-.2cm}
\resizebox{\linewidth}{!}{
\normalsize
\begin{tabular}{@{}cllcccccccccccc@{}}
	\toprule
	Configuration & RU, duplex, protocol && $M$ & $K$ & $f_\mathrm{c}$ [GHz] & $B$ [MHz] & $\Pmax$ [W] && $\alpha$ & $\gamma$ & $P_0$ [W]\textsuperscript{1} & $P_1$ [W]\textsuperscript{1} & $\Psleep$ [W] \\
	\midrule
	$\mathsf{4T4R}$ & RRU, FDD, 4G LTE 		&& $4$ & $2$ & $1.8$ & $20$ & $40$ 				&& $0.75$ & $5.33$ & \{$0$, $34.69$\} & \{$149.40$, $114.71$\} & $233.55$	\\
	$\mathsf{8T8R}$ & RRU, TDD, 5G NR 			&& $8$ & $4$ & $3.5$ & $100$ & $40$ 			&& $0.75$ & $5.38$ & \{$0$, $69.98$\} & \{$229.47$, $103.26$\} & $363.78$ 	\\
	$\mathsf{64T64R}$ & AAU, TDD, 5G NR 		&& $64$ & $8$ & $3.5$ & $100$ & $3.125$ 		&& $0.75$ & $3.50$ & \{$0$, $53.92$\} & \{$341.57$, $161.95$\} & $550.23$ 	\\
	\bottomrule
\end{tabular}
}
\\\vspace{.1cm}\small\textsuperscript{1}Left value is obtained with $\delta_\mathrm{\scriptscriptstyle{PA}}^\mathrm{dtx}=1$ and $\delta_\mathrm{\scriptscriptstyle{TRX}}^\mathrm{idle}=1$
as in~\cite{Golard_2024},
right value is obtained with $\delta_\mathrm{\scriptscriptstyle{PA}}^\mathrm{dtx}=0.25$ and $\delta_\mathrm{\scriptscriptstyle{TRX}}^\mathrm{idle}=0.5$.
\vspace{-.2cm}
\end{table*}

\section{Power Consumption Model}\label{section:power_model}

We adopt the model from~\cite{Golard_2024} that expresses the \gls{bs} power consumption averaged over the frame duration as

\begin{equation}\label{eq:P_cons_pimrc}
	\Pcons = \frac{\bar{P}_\mathrm{\scriptscriptstyle{PA}}}{\eta_\mathrm{s/c}^\mathrm{\scriptscriptstyle{PA}}}+
	\frac{\bar{P}_\mathrm{\scriptscriptstyle{AFE}}}{\eta_\mathrm{s/c}^\mathrm{\scriptscriptstyle{AFE}}}+
	\frac{\bar{P}_\mathrm{\scriptscriptstyle{DBB}}}{\eta_\mathrm{s/c}^\mathrm{\scriptscriptstyle{DBB}}}
\end{equation}
where $\bar{P}_\mathrm{\scriptscriptstyle{PA}}$, $\bar{P}_\mathrm{\scriptscriptstyle{AFE}}$, and $\bar{P}_\mathrm{\scriptscriptstyle{DBB}}$ are
the average power consumed by the \glspl{pa}, \gls{afe}, and \gls{dbb} across the frame duration (during both downlink and uplink), 
while $\eta_\mathrm{s/c}^\mathrm{\scriptscriptstyle{PA}}$,
$\eta_\mathrm{s/c}^\mathrm{\scriptscriptstyle{AFE}}$, and $\eta_\mathrm{s/c}^\mathrm{\scriptscriptstyle{DBB}}$
are the supplying and cooling efficiencies of the \glspl{pa}, \gls{afe}, and \gls{dbb}.
The above power consumption model predicts the consumption of sub-6 GHz 4G and 5G cellular \glspl{bs} in Belgium within 10-20\% uncertainty~\cite{Golard_2024}. 
It can be applied to \gls{ru} types including \gls{rru} and \gls{aau}, working either in \gls{fdd} or \gls{tdd}, and running either 4G \gls{lte} or 5G \gls{nr}.
Also, it is parametric such that parameter values can be changed to reflect different implementations of \gls{bs} components.
The model in~(\ref{eq:P_cons_pimrc}) can be adapted to the scenario that we target in this paper, and the 
average \gls{bs} consumed power across the frame can be expressed as a function of $\Na$, $\Ma$ and $\Pa$, giving
\begin{align}\label{eq:P_cons_general}
	\Pcons &= \frac{\Na}{N}\Ma\left(\frac{P_0}{M}+\gamma\Pa^\alpha\right) + \frac{\Ma}{M}P_1 + \Psleep
\end{align}
where $P_0$, $P_1$, $\Psleep$, $\gamma$ are non-negative parameters, and $\alpha\in[0.5,1]$. 
The term with $P_0$ depends on both the ratio of active time slots and active antennas.
The term depending on $\gamma\Pa^\alpha$ models the power consumed by the \glspl{pa}. 
The third term scales with the ratio of active antennas, while the last term is constant. 
Although~(\ref{eq:P_cons_general}) is kept compact to emphasize the dependence on the optimization variables,
we will show in the following that the model parameters depend on system parameters such as $K$, $\Pmax$, $B$ and $f_\mathrm{c}$, 
$\tau_\mathrm{\scriptscriptstyle{DL}}$ and $\tau_\mathrm{sig}$, etc.

To map the model in~(\ref{eq:P_cons_pimrc}) to~(\ref{eq:P_cons_general}), we recall that the average physical load in a frame is defined in~\cite{Golard_2024} 
as the time-average on all the spatial layers of the instantaneous load, which is given by ratio of the instantaneous number of data subcarriers to the total 
number of data subcarriers. When a time slot is active, we assume that all its data subcarriers are used to serve all the users using \gls{sdma}. Hence,
the average physical load in our scenario is given by $\Na/N$. Besides, we set the frame duration to $T_\mathrm{frame}$, the number of active (resp.~total) 
\glspl{pa} and TX/RX chains to $\Ma$ (resp.~$M$), and the number of active (resp.~total) spatial layers to $K_\mathrm{a}$ (resp.~$K$), 
where we consider $K_\mathrm{a}=K$. We also consider one cell, which corresponds to 
one sector and one frequency band. The full-load \gls{pa} output power and maximum \gls{pa} output power are set to $\Pmax$, 
which considers 8 dB output power back-off such that the \gls{pa} is in the linear regime.
We allow the \gls{pa} to operate at 
an arbitrary output power $\Pa$ during data transmission.
In the following, we provide the details on how this mapping is done and what are the different 
contributions to the total $\Pcons$.

\subsection{Power Amplifier}

The instantaneous power consumed by a \gls{pa} is modeled as the sum of a static consumption and a 
dynamic consumption that depends on the instantaneous output power $p$~\cite{Golard_2024}:
\begin{equation*}
	P_\mathrm{\scriptscriptstyle{PA}}(p) = \xi\frac{P_\mathrm{max}^\alpha}{\eta_\mathrm{\scriptscriptstyle{PA}}^\mathrm{max}}+
	(1-\xi)\frac{P_\mathrm{max}^{1-\alpha}p^\alpha}{\eta_\mathrm{\scriptscriptstyle{PA}}^\mathrm{max}}
\end{equation*}
where $\xi\in[0,1]$ and $\eta_\mathrm{\scriptscriptstyle{PA}}^\mathrm{max}$ is the maximum \gls{pa} efficiency~\cite[eq.~(5)]{Golard_2024} 
that depends on the \gls{pa} technology, on the carrier frequency $f_\mathrm{c}$ and on $\Pmax$.
Parameters of common \gls{pa} architectures are in~\cite[Table I]{Golard_2024}.
Each of the $\Ma$ active \glspl{pa} can be in one of four states (see Fig.~\ref{fig:frame}):
\begin{itemize}[leftmargin=*]
	\item Downlink mode and reference signal transmission, where $p=\zeta_\mathrm{sig}\Pmax$, with $\zeta_\mathrm{sig}\in(0,1)$ being the ratio of reference signaling power to $\Pmax$.
	This gives a power consumption equal to $\tau_\mathrm{\scriptscriptstyle{DL}}\tau_\mathrm{sig}P_\mathrm{\scriptscriptstyle{PA}}(\zeta_\mathrm{sig}\Pmax)$.
	\item Downlink mode and data transmission, with $p=\Pa$, resulting in a power consumption of 
	$\tau_\mathrm{\scriptscriptstyle{DL}}(1-\tau_\mathrm{sig})\frac{\Na}{N}P_\mathrm{\scriptscriptstyle{PA}}(\Pa)$.
	\item Downlink mode and no data and no reference signal transmission, with $p=0$. This corresponds to \gls{dtx} mode. The power consumption is
	$\tau_\mathrm{\scriptscriptstyle{DL}}(1-\tau_\mathrm{sig})\left(1-\frac{\Na}{N}\right)P_\mathrm{\scriptscriptstyle{PA}}(0)\delta_\mathrm{\scriptscriptstyle{PA}}^\mathrm{dtx}$,
	where $\delta_\mathrm{\scriptscriptstyle{PA}}^\mathrm{dtx}\in[0,1]$ is a reduction factor when in \gls{dtx}.
	\item Idle state when not in downlink mode, with $p=0$, resulting in a power consumption given by 
	$(1-\tau_\mathrm{\scriptscriptstyle{DL}})P_\mathrm{\scriptscriptstyle{PA}}(0)\delta_\mathrm{\scriptscriptstyle{PA}}^\mathrm{idle}$, 
	where $\delta_\mathrm{\scriptscriptstyle{PA}}^\mathrm{idle}\in[0,1]$ is a reduction factor when in idle mode.
\end{itemize}
The $M-\Ma$ non-active \glspl{pa} are instead in sleep mode and their consumed power is 
$(M-\Ma)P_\mathrm{\scriptscriptstyle{PA}}(0)\delta_\mathrm{\scriptscriptstyle{PA}}^\mathrm{sleep}$,
where $\delta_\mathrm{\scriptscriptstyle{PA}}^\mathrm{sleep}\in[0,1]$ is a reduction factor when in sleep mode.
The reduction factors can represent opposite situations, \eg, $\delta_\mathrm{\scriptscriptstyle{PA}}^\mathrm{dtx}=1$ (no \gls{mudtx}) and $\delta_\mathrm{\scriptscriptstyle{PA}}^\mathrm{dtx}=0$ (ideal \gls{mudtx}).
By grouping the above terms, the consumed power by the \glspl{pa} averaged over the frame duration is given by~\cite[eq.~(6)]{Golard_2024}
\begin{align}
	\bar{P}_\mathrm{\scriptscriptstyle{PA}} &= 
	\frac{\Na}{N}\Ma\tau_\mathrm{\scriptscriptstyle{DL}}\left(1-\tau_\mathrm{sig}\right)
	\Big(P_\mathrm{{\scriptscriptstyle{PA}}}\left(\Pa\right)-
	P_\mathrm{{\scriptscriptstyle{PA}}}(0)\delta_\mathrm{\scriptscriptstyle{PA}}^\mathrm{dtx}\Big) \nonumber\\
	&+ \Ma\bar{P}_\mathrm{\scriptscriptstyle{PA}\scriptstyle{,1}}+MP_\mathrm{\scriptscriptstyle{PA}}(0)\delta_\mathrm{\scriptscriptstyle{PA}}^\mathrm{sleep} \nonumber
\end{align}
where we defined 
\begin{align}
	\bar{P}_\mathrm{\scriptscriptstyle{PA}\scriptstyle{,1}} &\triangleq \tau_\mathrm{\scriptscriptstyle{DL}}\tau_\mathrm{sig}
	P_\mathrm{{\scriptscriptstyle{PA}}}\left(\zeta_\mathrm{sig}\Pmax\right)+
	\tau_\mathrm{\scriptscriptstyle{DL}}\left(1-\tau_\mathrm{sig}\right)
	P_\mathrm{{\scriptscriptstyle{PA}}}(0)\delta_\mathrm{\scriptscriptstyle{PA}}^\mathrm{dtx} \nonumber \\
	&+ \left(1-\tau_\mathrm{\scriptscriptstyle{DL}}\right)P_\mathrm{\scriptscriptstyle{PA}}(0)
	\delta_\mathrm{\scriptscriptstyle{PA}}^\mathrm{idle}-
	P_\mathrm{\scriptscriptstyle{PA}}(0)\delta_\mathrm{\scriptscriptstyle{PA}}^\mathrm{sleep}. \label{eq:P_PA,2_bar}
\end{align}
We can therefore identify in $\bar{P}_\mathrm{\scriptscriptstyle{PA}}$ the terms composing $\gamma$ and $P_0/M$,
as well as the terms that are part of $P_1/M$ and $\Psleep$.

\subsection{Analog Front-End}

The \gls{afe} performs digital-to-analog/analog-to-digital conversions, up-/down-conversion, pre-driving, low-noise amplification, frequency synthesis and general control.
We consider the miscellaneous functions to be always active and the number of TX and RX chains to be the same.
The $\Ma$ active TX/RX chains can be in (i) working mode, where we assume that the same power is consumed in downlink and uplink, 
and (ii) idle mode. The $M-\Ma$ non-active TX/RX chains are in sleep mode instead. 
By using~\cite[eq.~(10)]{Golard_2024}, the consumed power by the \gls{afe} averaged over the frame duration is expressed as
\begin{equation*}
	\bar{P}_\mathrm{\scriptscriptstyle{AFE}}
	= \Ma\bar{P}_\mathrm{\scriptscriptstyle{AFE}\scriptstyle{,1}}+
	\bar{P}_\mathrm{\scriptscriptstyle{AFE}\scriptstyle{,sleep}}
\end{equation*}
where we defined
\begin{align}
	\bar{P}_\mathrm{\scriptscriptstyle{AFE}\scriptstyle{,1}} &\triangleq P_\mathrm{\scriptscriptstyle{TRX}}\big(\left(\tau_\mathrm{\scriptscriptstyle{DL}}+
	\tau_\mathrm{\scriptscriptstyle{UL}}\right) +\left(2-\tau_\mathrm{\scriptscriptstyle{DL}} -
	\tau_\mathrm{\scriptscriptstyle{UL}}\right)\delta_\mathrm{\scriptscriptstyle{TRX}}^\mathrm{idle}
	\nonumber\\ &-2\delta_\mathrm{\scriptscriptstyle{TRX}}^\mathrm{sleep}\big) \label{eq:P_AFE,1_bar}\\
	\bar{P}_\mathrm{\scriptscriptstyle{AFE}\scriptstyle{,sleep}} &\triangleq 
	P_\mathrm{misc}+2MP_\mathrm{\scriptscriptstyle{TRX}}\delta_\mathrm{\scriptscriptstyle{TRX}}^\mathrm{sleep} \label{eq:P_AFE,sleep_bar}
\end{align}
with $P_\mathrm{misc}$ scaling with the bandwidth $B$ and $M\Pmax$, $P_\mathrm{\scriptscriptstyle{TRX}}$ 
scaling with $B$ and $\Pmax$~\cite[eq.~(9) and~(10)]{Golard_2024}, and 
$\delta_\mathrm{\scriptscriptstyle{TRX}}^\mathrm{idle}, \delta_\mathrm{\scriptscriptstyle{TRX}}^\mathrm{sleep} \in[0,1]$ representing the
reduction factors when the \gls{afe} is in idle and sleep modes.
The \gls{afe} contribution is thus (physical) load-independent and part of $P_1/M$ and $\Psleep$.

\subsection{Digital Baseband}

The operations performed by the \gls{dbb} are physical resource scheduling, (de)coding, (de)modulation, digital precoding and decoding, etc.
The \gls{dbb} consumption is modeled as the consumption at the data link layer $P_\mathrm{link}^\mathrm{ref}$, which is considered constant, plus the consumption at the physical layer,
which comprises either working or idle mode. The average \gls{dbb} consumed power over the frame is then given by~\cite[eq.~(11)]{Golard_2024}
\begin{equation}\label{eq:P_DBB_bar}
\begin{aligned}
	\bar{P}_\mathrm{\scriptscriptstyle{DBB}} = P_\mathrm{link}^\mathrm{ref}+\tilde{P}_\mathrm{phy}^\mathrm{ref}
	\big(\left(\tau_\mathrm{\scriptscriptstyle{DL}}+
	\tau_\mathrm{\scriptscriptstyle{UL}}\right) + \left(2-\tau_\mathrm{\scriptscriptstyle{DL}}-
	\tau_\mathrm{\scriptscriptstyle{UL}}\right)\delta_\mathrm{phy}^\mathrm{idle}\big)
\end{aligned}
\end{equation}
where $P_\mathrm{link}^\mathrm{ref}$ is a constant and $\tilde{P}_\mathrm{phy}^\mathrm{ref}$ is linearly proportional to the number of users $K$ and scales with $B$.
As for the \gls{afe}, the \gls{dbb} consumption is (physical) load-independent and this time is only part of $\Psleep$.

\subsection{Power Supply and Cooling Systems}
Power converters (AC/DC and DC/DC) and active cooling modules in a \gls{bs} also contribute to the \gls{bs} consumption. We consider that the first introduce an overhead with respect to the total consumption
with an efficiency $\eta_\mathrm{supply}$, while the second introduce component-wise overheads with efficiencies $\eta_\mathrm{cool}^\mathrm{\scriptscriptstyle{PA}},
\eta_\mathrm{cool}^\mathrm{\scriptscriptstyle{AFE}}, \eta_\mathrm{cool}^\mathrm{\scriptscriptstyle{DBB}}\in (0,1]$~\cite{Golard_2024}. In case of no active cooling, 
$\eta_\mathrm{cool}^\mathrm{\scriptscriptstyle{PA}}=\eta_\mathrm{cool}^\mathrm{\scriptscriptstyle{AFE}}=\eta_\mathrm{cool}^\mathrm{\scriptscriptstyle{DBB}}=1$.
The supply and cooling efficiencies are eventually defined as $\eta_\mathrm{s/c}^\mathrm{\scriptscriptstyle{PA}}=\eta_\mathrm{supply}\eta_\mathrm{cool}^\mathrm{\scriptscriptstyle{PA}}$,
$\eta_\mathrm{s/c}^\mathrm{\scriptscriptstyle{AFE}}=\eta_\mathrm{supply}\eta_\mathrm{cool}^\mathrm{\scriptscriptstyle{AFE}}$, 
and $\eta_\mathrm{s/c}^\mathrm{\scriptscriptstyle{DBB}}=\eta_\mathrm{supply}\eta_\mathrm{cool}^\mathrm{\scriptscriptstyle{DBB}}$.
Now, using~(\ref{eq:P_PA,2_bar}),~(\ref{eq:P_AFE,1_bar}),~(\ref{eq:P_AFE,sleep_bar}),~(\ref{eq:P_DBB_bar}) and linking them to~(\ref{eq:P_cons_pimrc})
we can express the parameters of our power consumption model~(\ref{eq:P_cons_general}) as follows
\begin{equation*}
\begin{aligned}
	\gamma &= \frac{\tau_\mathrm{\scriptscriptstyle{DL}}\left(1-\tau_\mathrm{sig}\right)}{\eta_\mathrm{s/c}^\mathrm{\scriptscriptstyle{PA}}}
	(1-\xi)\frac{P_\mathrm{max}^{1-\alpha}}{\eta_\mathrm{\scriptscriptstyle{PA}}^\mathrm{max}}\\
	\frac{P_0}{M} &= \frac{\tau_\mathrm{\scriptscriptstyle{DL}}(1-\tau_\mathrm{sig})}{\eta_\mathrm{s/c}^\mathrm{\scriptscriptstyle{PA}}}
	\xi\frac{P_\mathrm{max}^\alpha}{\eta_\mathrm{\scriptscriptstyle{PA}}^\mathrm{max}}
	\big(1-\delta_\mathrm{\scriptscriptstyle{PA}}^\mathrm{dtx}\big) \\
	\frac{P_1}{M} &= \frac{\bar{P}_\mathrm{\scriptscriptstyle{PA}\scriptstyle{,1}}}{\eta_\mathrm{s/c}^\mathrm{\scriptscriptstyle{PA}}}
	+\frac{\bar{P}_\mathrm{\scriptscriptstyle{AFE}\scriptstyle{,1}}}{\eta_\mathrm{s/c}^\mathrm{\scriptscriptstyle{AFE}}}
\end{aligned}
\end{equation*}
\begin{equation*}
\begin{aligned}
	\Psleep &= \frac{M \delta_\mathrm{\scriptscriptstyle{PA}}^\mathrm{sleep}}{\eta_\mathrm{s/c}^\mathrm{\scriptscriptstyle{PA}}}
	\xi\frac{P_\mathrm{max}^\alpha}{\eta_\mathrm{\scriptscriptstyle{PA}}^\mathrm{max}}+
	\frac{\bar{P}_\mathrm{\scriptscriptstyle{AFE}\scriptstyle{,sleep}}}{\eta_\mathrm{s/c}^\mathrm{\scriptscriptstyle{AFE}}}
	+ \frac{\bar{P}_\mathrm{\scriptscriptstyle{DBB}}}{\eta_\mathrm{s/c}^\mathrm{\scriptscriptstyle{DBB}}}.
\end{aligned}
\end{equation*}
To compute the above parameters, we select three \gls{bs} configurations ($\mathsf{64T64R}$, $\mathsf{8T8R}$, and $\mathsf{4T4R}$) and utilize the values in~\cite[Table II]{Golard_2024}. 
We consider two pairs of values for the reduction factors $\delta_\mathrm{\scriptscriptstyle{PA}}^\mathrm{dtx}$ and $\delta_\mathrm{\scriptscriptstyle{TRX}}^\mathrm{idle}$,
leading to two cases:
\begin{itemize}[leftmargin=*]
\item \underline{Disabled time-domain hardware power-saving modes}, with $\delta_\mathrm{\scriptscriptstyle{PA}}^\mathrm{dtx}=1$ and $\delta_\mathrm{\scriptscriptstyle{TRX}}^\mathrm{idle}=1$ as considered in~\cite{Golard_2024}.
This corresponds to not using PA \textmu DTX and \gls{afe} idle modes.
\item \underline{Enabled time-domain hardware power-saving modes}, with $\delta_\mathrm{\scriptscriptstyle{PA}}^\mathrm{dtx}=0.25$ and $\delta_\mathrm{\scriptscriptstyle{TRX}}^\mathrm{idle}=0.5$.
The obtained model parameters reflect the \glspl{bs} under study in~\cite{Golard_2024} but utilizing related to \gls{pa} \gls{mudtx} and \gls{afe} idle modes.
The values of $\delta_\mathrm{\scriptscriptstyle{PA}}^\mathrm{dtx}$ and $\delta_\mathrm{\scriptscriptstyle{TRX}}^\mathrm{idle}$ are set to be equal to $\delta_\mathrm{\scriptscriptstyle{PA}}^\mathrm{idle}$
and $\delta_\mathrm{\scriptscriptstyle{TRX}}^\mathrm{sleep}$, respectively.
\end{itemize}
The obtained values of~(\ref{eq:P_cons_general}) are given in Table~\ref{tab:parameters_Pcons}.

\section{Optimal Allocation of Power, Time and Spatial Resources}\label{section:optimal_allocation}

This section considers the problem
\begin{align}\label{eq:general_opt_prob}
	\min_{N_\mathrm{a}, M_{\mathrm{a}},\Pa} P_{\mathrm{cons}}
	\ \mathrm{s.t.} \ R_k=\frac{N_{\mathrm{a}}}{N}\log_2\left(1+\frac{p_{k}}{\sigma_k^2}\right),\ k=1,\dotsc,K.
\end{align}
\begin{figure*}[!t]
	\begin{align}\label{eq:simplified_prob2}
		\min_{\substack{
		0 \leq N_\mathrm{a}\leq N\\ K\leq \Ma \leq M}}\ P_{\mathrm{cons}}=\frac{\Na}{N}\frac{\Ma}{M}P_{0}+\frac{\Na}{N}\Ma\gamma
		\left(\frac{1}{\Ma(\Ma-K)}\sum_{k=1}^{K}\frac{\sigma_k^2}{\beta_k}\left(2^{R_k\frac{N}{\Na}}-1\right)\right)^{\alpha}
		+\frac{\Ma}{M}P_1 +\Psleep
		\quad\mathrm{s.t.\ }  \left(\mathcal{C}_{P_\mathrm{max}}\right)
	\end{align}
	\vspace*{-15pt}
	\hrulefill
\end{figure*}

\noindent
In words, we consider the minimization of $P_{\mathrm{cons}}$ by optimizing the number of active time slots $\Na$ and active antennas $\Ma$, and the transmit power at active antennas $\Pa$. 
The optimization is subject to a target rate constraint $R_k$ for each user. Moreover, as detailed in Section~\ref{section:transmission_model}, the optimization variables have additional constraints 
(not directly shown in (\ref{eq:general_opt_prob})) given their physical meaning: $0 \leq N_\mathrm{a}\leq N$, $K\leq \Ma \leq M$ and $0 \leq \Pa \leq \Pmax$, $\Na$ and $\Ma$ are integers while $\Pa$ is continuous.
This section adopts a step-by-step approach to solve problem~(\ref{eq:general_opt_prob}) and shows that it can be efficiently solved through a two-dimensional differentiable convex problem. 
More specifically, in Section~\ref{subsection:optimal_P_a}, we first give the optimal transmit power $\Pa$ as a function of $\Na$ and $\Ma$. 
Then in Section~\ref{subsection:optimal_N_a_M_a}, we find the optimal time and spatial resources $\Na$ and $\Ma$ to allocate.  
Section~\ref{subsection:approx_solution} proposes an approximation of the optimal solution.

\subsection{Optimal Transmit Power at Active Antennas}\label{subsection:optimal_P_a}

As a first step to solve problem~(\ref{eq:general_opt_prob}), we can find that the active antenna transmit power $\Pa$ can be expressed in terms of $\Na$ and $\Ma$ through the user rate constraint. 
Indeed, we can solve for $p_k$
\begin{align}\label{eq:R_k}
	\!\!R_k&=\frac{\Na}{N}\log_2\left(1+\frac{p_{k}}{\sigma_k^2}\right) \iff p_k=\sigma_k^2 \left(2^{R_k\frac{N}{\Na}}-1\right)
\end{align}
and substitute the expression of $p_k$ into~(\ref{eq:general_Pa_expression}) giving
\begin{align}\label{eq:general_Pa}
	P_{\mathrm{a}}=\frac{1}{M_{\mathrm{a}}(M_{\mathrm{a}}-K)}\sum_{k=1}^{K}\frac{\sigma_k^2}{\beta_k} \left(2^{R_k\frac{N}{\Na}}-1\right).
\end{align}
The term $R_k\frac{N}{\Na}$ can be interpreted as the user rate per active time slot, where $R_k$ gives the frame-averaged number of bits per subcarrier and per time slot.
Substituting~(\ref{eq:general_Pa}) into problem~(\ref{eq:general_opt_prob}), 
it simplifies to the two-dimensional problem~(\ref{eq:simplified_prob2}) in $\Ma$ and $\Na$. Note that the user rate constraints are implicitly taken into account and thus satisfied. 
Moreover, the maximal power constraint $\Pa\leq \Pmax$ can be converted in terms of $\Ma$ and $\Na$ to
\begin{align}\label{eq:P_max_constraint}
	\!\left(\mathcal{C}_{P_\mathrm{max}}\right)\!: \frac{1}{M_{\mathrm{a}}(M_{\mathrm{a}}-K)}\sum_{k=1}^{K}\frac{\sigma_k^2}{\beta_k} \left(2^{R_k\frac{N}{\Na}}-1\right)\leq \Pmax.
\end{align}
Clearly, $\Pa$ is minimized when $\Ma=M$ and $\Na=N$ leading to the definition
\begin{align}\label{eq:P_a_min}
	\Pamin=\frac{1}{M(M-K)}\sum_{k=1}^{K}\frac{\sigma_k^2}{\beta_k} \left(2^{R_k}-1\right).
\end{align}
Hence, problem~(\ref{eq:general_opt_prob}) is feasible if $\Pamin \leq \Pmax$. 
In other words, activating all resources ($\Ma=M$, $\Na=N$) allows the system to meet the target users' rates within the power budget. 
As shown in Appendix~\ref{appendix:P_max_simplified}, the inequality~(\ref{eq:P_max_constraint}) can be further simplified as
\begin{align}\label{eq:P_max_constraint_simplified}
	\Ma \geq \frac{K}{2}+\frac{1}{2}\sqrt{K^2+4\sum_{k=1}^{K}\rho_k^{-1}\left(2^{R_k\frac{N}{\Na}}-1\right) }
\end{align}
where $\rho_k=\Pmax\beta_k/\sigma_k^2$ and which implies $\Ma\geq K$. Two useful quantities can be derived from this inequality. Firstly, the minimum amount of active antennas, defined as $M_\mathrm{a,min}$, 
is obtained by finding the minimum value of $\Ma$ that satisfies~(\ref{eq:P_max_constraint_simplified}) when $\Na=N$
\begin{align} \label{eq:M_a_min}
	M_\mathrm{a,min} = \left\lceil\frac{K}{2}+\frac{1}{2}\sqrt{K^2+4\sum_{k=1}^{K}\rho_k^{-1}\left(2^{R_k}-1\right) }\;\right\rceil.
\end{align}
Secondly, the minimum amount of active slots, defined as $N_\mathrm{a,min}$, is obtained by finding the minimum value of $\Na$ that satisfies~(\ref{eq:P_max_constraint_simplified}) when $\Ma=M$
\begin{align} \label{eq:N_a_min}
	M \geq \frac{K}{2}+\frac{1}{2}\sqrt{K^2+4\sum_{k=1}^{K}\rho_k^{-1}\left(2^{R_k\frac{N}{\Namin}}-1\right) }.
\end{align}
In general, $N_\mathrm{a,min}$ does not have a closed-form solution.

\vspace{-.25cm}
\subsection{Optimal Space and Time Resources}\label{subsection:optimal_N_a_M_a}
Let us consider a continuous relaxation of problem~(\ref{eq:simplified_prob2}). Making the change of variable $x=N/\Na$, $y=\Ma$ and removing the constant term $\Psleep$ not impacting the optimization, 
the cost function becomes
\begin{align}\label{eq:f_xy}
	f(x,y)&=\frac{y}{x} \frac{P_{0}}{M}+\frac{y}{x}\gamma\left(\frac{1}{y(y-K)}\sum_{k=1}^{K}\frac{\sigma_k^2}{\beta_k} \left(2^{R_k x }-1\right)\right)^{\alpha}\nonumber\\ 
	&+\frac{y}{M}P_1.
\end{align}
In Appendix~\ref{appendix:convexity_f}, we show that $f(x,y)$ is
convex in $(x,y)$.\footnote{As a note, 
defining $x$ as $\Na$ or $\Na/N$ instead of $N/\Na$ would not have led to a convex function $f(x,y)$.} 

\noindent Moreover, the constraints $0 \leq N_\mathrm{a}\leq N$,
$K\leq\Ma \leq M$
and~(\ref{eq:P_max_constraint_simplified}) imply that $x$ and $y$ must belong to the feasibility domain $\mathcal{D}$ defined as
\begin{align}
	\mathcal{D}=\Big\{&1\leq x, y\leq M,\nonumber\\
	&y \geq \frac{K}{2}+\frac{1}{2}\sqrt{K^2+4\sum_{k=1}^{K}\rho_k^{-1}\left(2^{R_kx}-1\right)}\;\Big\}\label{eq:set_D}.
\end{align}
In a convex problem, both the objective function and the constraints need to be convex. 
In Appendix~\ref{appendix:convexity_D}, we show that a sufficient condition for the convexity of the set $\mathcal{D}$ is
\begin{align}\label{eq:convexity_constraint}
	\rho_k=\frac{\Pmax \beta_k}{\sigma_k^2}\geq \frac{2}{K}\ \text{for } k=1,\dotsc,K.
\end{align}
This can be seen as a minimal average \gls{snr} constraint at each user when using a single transmit antenna at maximal power. 
This is a sufficient condition, which is required to make a strong general statement but can be neglected in certain regimes. 
For instance, for a small number of users and/or small target data rates, the $\Pmax$ constraint
is not binding and~(\ref{eq:convexity_constraint}) can be left out. In the extreme opposite case, for a large number of users and/or large target data rates, 
a sufficiently large \gls{snr} at each user will be necessary to make the problem feasible, implying that~(\ref{eq:convexity_constraint}) will be satisfied. 

\begin{figure}[!t]
	\centering
	\resizebox{.92\linewidth}{!}{
		\includegraphics{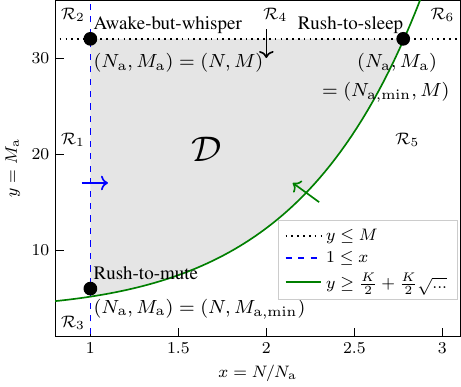}
	}
	\vspace{-.3cm}
	\caption{Example of feasibility domain $\mathcal{D}$ for $M=32$, $K=4$, $\Pmax=1$, $\beta_k=1$, $\sigma_k^2=10^{-1}$, $R_k=4$, for $k=1,\dotsc,K$.
	The point $(N,\Mamin)$ has a small offset from the minimum $y$ due to the ceiling operation in~(\ref{eq:M_a_min}).}
	\label{fig:set_D}
	\vspace{-.2cm}
\end{figure}

An illustration of~$\mathcal{D}$ is shown in Fig.~\ref{fig:set_D}, which resembles a triangle with a curved bottom right side. 
Each of the three vertices corresponds to a specific energy-saving regime:
\begin{itemize}[leftmargin=*]
	\item \textbf{Rush-to-sleep} (time-domain only): all spatial resources are active with full transmit power giving $\Ma=M$, $\Pa=\Pmax$ with minimum active time slots $\Na=\Namin$.
	\item \textbf{Rush-to-mute} (spatial-domain only): all time slots are active with full transmit power giving $\Na=N$, $\Pa=\Pmax$ with minimum number of active antennas $\Ma=\Mamin$.
	\item \textbf{Awake-but-whisper} (power-domain only): all spatial and temporal resources are activated giving $x=1$, $\Na=N$ and $\Ma=M$ with minimum transmit power $\Pamin$.
\end{itemize}
The awake-but-whisper and rush-to-mute approaches have a zero \gls{pa} idle-mode duration while the rush-to-sleep has the largest one, which also minimizes latency.
As a result of the convexity of $f(x,y)$ and $\mathcal{D}$, the problem
\begin{align}\label{eq:convex_problem}
	(\bar{x},\bar{y})=\arg \min_{(x,y)\in \mathcal{D}} \ f(x,y)
\end{align}
is convex and differentiable. We can thus use simple and efficient techniques to perform the minimization, with guarantee to find the global minimum.
Another favourable consequence of the convexity of~(\ref{eq:convex_problem}) is that the solution of the original (discrete) problem can be found through the ceil-floor operator
\begin{align}\label{eq:optimal_Na_Ma}
	(\Na,\Ma)=\big\lfloor N/\bar{x},\bar{y}\big\rceil.
\end{align}
Indeed, for a convex problem, the optimal discrete point should be one of the nearest (feasible) neighbours.

\subsection{Approximation of Optimal Solution}\label{subsection:approx_solution}
The ceil-floor operation requires to evaluate and compare $P_{\mathrm{cons}}$ for four possible values as $\Na$ is either $\ceil{N/\bar{x}}$ or $\floor{N/\bar{x}}$ 
while $\Ma$ is either $\floor{\bar{y}}$ or $\ceil{\bar{y}}$. In practice, if $N$ is large, the ceil-floor operator can be approximated by a simpler rounding operation $\Na=[N/\bar{x}]$, 
not requiring any comparison (but just checking the feasibility). The resulting deviation from the optimal value of the cost function vanishes as $\mathcal{O}(1/N)$. 
The intuition is that $P_{\mathrm{cons}}$ depends on $\Na$ through the ratio $\Na/N$, implying that the discretization error is at most $1/N$ and that asymptotically, 
the ratio $\Na/N$ can take an infinite number (not only discrete) number of points between $0$ and~$1$. A similar reasoning can be applied for the number of antennas: replacing the ceil-floor 
operation by a rounding operator $\Ma=[\bar{y}]$. The resulting approximation error decays as $\mathcal{O}(1/\Ma^{2\alpha})$ as $\Ma$ grows large. 
In summary, as shown in Appendix~\ref{appendix:ceil_floor}, using
\begin{align}\label{eq:sol_with_round}
	\Na=\round{N/\bar{x}},\ \Ma= \round{\bar{y}}
\end{align}
provides the minimal $P_{\mathrm{cons}}$ up to an error term that decreases as $\mathcal{O}\big(1/N+1/M+1/\Ma^{2\alpha}\big)$ when the number of time slots
$N\rightarrow +\infty$ and the number of active antennas $\Ma\rightarrow +\infty$. 
In practice, the large $N$ assumption is realistic as frames are usually made of many symbols. The impact of the second and third contributions to the 
error term depends on which \gls{bs} configuration is selected because that fixes the number of antennas $M$ and users $K$, and
imposes a constraint $\Ma\geq K$.

As a final remark, problem~(\ref{eq:simplified_prob2}) only depends on the large-scale fading coefficients of the user-to-\gls{bs} channels, not on the instantaneous channels. 
These coefficients are simpler to estimate than the instantaneous channels as they vary more slowly in time. Moreover, the optimal allocation of time-space-power resources should only be done
once per channel stationary time, \ie, when the large-scale fading coefficients change, or when a different scheduling policy should be found. 
The optimization is thus already light given the problem convexity, but on top, it has to be updated on a relatively slow time scale.

\subsection{Practical Algorithm and Complexity}

To solve problem~(\ref{eq:convex_problem}), we utilize Algorithm~\ref{alg1} where every
unconstrained problem is solved via Newton's method~\cite[Algorithm 9.5]{boyd_convex_2017} with tolerance of $10^{-8}$. 
After checking the problem feasibility, we solve the unconstrained two-dimensional problem $\min_{x,y} f(x,y)$.
Through simulations, we observe that from $20$ to $30$ iterations are necessary to reach convergence.
The gradient and Hessian of $f(x,y)$ depend on combinations of powers of 
$\phi(x)=\sum_{k=1}^K\left(\sigma_k^2/\beta_k\right)\left(2^{R_kx}-1\right)$, 
$\phi'(x)$, $\phi''(x)$, $x$ and $y$, where their values can be when reused among
different terms in order to reduce complexity.
If the solution of the problem lies in $\mathcal{D}$, the constraints are satisfied and the problem is solved. 
Otherwise, looking at the 
geometry of $\mathcal{D}$, either one or two constraints are not satisfied. We can identify six cases corresponding so six domains
$\mathcal{R}_i$, $i=1,\dotsc,6$, illustrated also in Fig.~\ref{fig:set_D}. Let us define 
\begin{equation*}
	y_\mathrm{min}(x)=\frac{K}{2}+\frac{1}{2}\sqrt{K^2+4\sum_{k=1}^{K}\rho_k^{-1}\left(2^{R_kx}-1\right)}
\end{equation*}
as the value of $y$ that satisfies the power constraint with equality. 
The feasibility domain can therefore be written as $\mathcal{D}=\{1\leq x, y\leq M, y\geq y_\mathrm{min}(x)\}$.
When the first, second and third constraints are not satisfied we obtain the domains
$\mathcal{R}_1=\{1>x, y\leq M, y\geq y_\mathrm{min}(x)\}$,
$\mathcal{R}_4=\{1\leq x, y> M, y\geq y_\mathrm{min}(x)\}$ and
$\mathcal{R}_5=\{1\leq x, y\leq M, y< y_\mathrm{min}(x)\}$, respectively.
When only the third, second and first constraints are satisfied we obtain the domains
$\mathcal{R}_2=\{1>x, y> M, y\geq y_\mathrm{min}(x)\}$,
$\mathcal{R}_3=\{1>x, y\leq M, y< y_\mathrm{min}(x)\}$ and
$\mathcal{R}_6=\{1\leq x, y> M, y< y_\mathrm{min}(x)\}$, respectively.
If the solution of $\min_{x,y} f(x,y)$ lies in $\mathcal{R}_1$, the first constraint is not satisfied. 
Making it binding sets the value of $x$ to $1$, and we can then solve the one-dimensional unconstrained problem 
$\min_y f(1,y)$. If the obtained solution is either larger than $M$ or smaller than $y_\mathrm{min}(1)$, we make 
the unsatisfied constraint binding and this gives the final value of $y$. In case the solution of
$\min_{x,y} f(x,y)$ lies in $\mathcal{R}_2$, we let the first constraint bind fixing $x$ to $1$, solve the 
problem $\min_y f(1,y)$ and further check the other two constraints in order to make the obtained $y$ within bounds. 
We also let the second constraint bind fixing $y$ to $M$, solve the problem $\min_x f(x,M)$ and further check the other two constraints
fixing $x$ either to $1$ or to $x_\mathrm{max}$ if not satisfied, where $x_\mathrm{max}$ is the value solving 
$y_\mathrm{min}(x_\mathrm{max})=M$. We then select among the two solutions, the one that minimizes $f(x,y)$. 
The same procedure is applied in the remaining four cases. Therefore, we might need to solve two of the three one-dimensional unconstrained problems 
$\min_y f(1,y)$, $\min_x f(x,M)$ and $\min_x f(x,y_\mathrm{min}(x))$. Through simulations, we observe that $5$ to $20$ iterations are necessary to solve 
any of the three problems. Computing the first and second derivatives of the three functions involves similar computations to the ones
required for the gradient and Hessian of $f(x,y)$ but with fewer terms given that the problems are one-dimensional.

\begin{algorithm}[t!]
\small
\caption{Unconstrained optimization with inspection of the constraints to solve problem~(\ref{eq:convex_problem}) and retrieve~(\ref{eq:sol_with_round})}
\small
\begin{minipage}{\linewidth}
\setstretch{1.1}
\begin{algorithmic}[1]
	\State \textbf{Input:} $\{R_k\}$, $\{\beta_k\}$, $\{\sigma_k^2\}$, $M$, $N$, $K$, $\gamma$, $\alpha$, $P_0$, $P_1$, $\Pmax$
	\State Compute $P_\mathrm{a,min}$ via~(\ref{eq:P_a_min}). If $P_\mathrm{a,min} > \Pmax$, quit
	\State Set $(\bar{x},\bar{y})\gets\argmin_{(x,y)} f(x,y)$
	\If{$(\bar{x},\bar{y})\in\mathcal{D}$}
		\State Break, problem solved
	\ElsIf{$(\bar{x},\bar{y})\in\mathcal{R}_1$}
		\State Set $\bar{x}\gets1$, $\bar{y} \gets \argmin_{y} f(1,y)$ \par and
		make $\bar{y}$ within bounds
	\ElsIf{$(\bar{x},\bar{y})\in\mathcal{R}_2$}
		\State Set $\bar{x}_1\gets1$, $\bar{y}_1\gets \argmin_{y} f(1,y)$ \par and
		make $\bar{y}_1$ within bounds
		\State Set $\bar{y}_2\gets M$, $\bar{x}_2\gets \argmin_{x} f(x,M)$ \par and 
		make $\bar{x}_2$ within bounds
		\State Set $(\bar{x},\bar{y})\gets$ best\textsuperscript{$\star$} among $(\bar{x}_1,\bar{y}_1)$ and $(\bar{x}_2,\bar{y}_2)$
	\ElsIf{$(\bar{x},\bar{y})\in\mathcal{R}_3$}
		\State Set $\bar{x}_1\gets1$, $\bar{y}_1\gets \argmin_{y} f(1,y)$ \par and 
		make $\bar{y}_1$ within bounds
		\State Set $\bar{x}_2\gets \argmin_{x} f(x,y_\mathrm{min}(x))$, make $\bar{x}_2$ 
		within bounds \par and set $\bar{y}_2\gets y_\mathrm{min}(\bar{x}_2)$
		\State Set $(\bar{x},\bar{y})\gets$ best\textsuperscript{$\star$} among $(\bar{x}_1,\bar{y}_1)$ and $(\bar{x}_2,\bar{y}_2)$
	\ElsIf{$(\bar{x},\bar{y})\in\mathcal{R}_4$}
		\State Set $\bar{y}\gets M$, $\bar{x}\gets \argmin_{x} f(x,M)$ \par and 
		make $\bar{x}$ within bounds
	\ElsIf{$(\bar{x},\bar{y})\in\mathcal{R}_5$}
		\State Set $\bar{x}\gets\argmin_{x} f(x,y_\mathrm{min}(x))$, make $\bar{x}$ 
		within bounds \par and set $\bar{y}\gets y_\mathrm{min}(\bar{x})$
	\ElsIf{$(\bar{x},\bar{y})\in\mathcal{R}_6$}
		\State Set $\bar{y}_1\gets M$, $\bar{x}_1\gets\argmin_{x} f(x,M)$
		\par and make $\bar{x}_1$ within bounds
		\State Set $\bar{x}_2\gets\argmin_{x} f(x,y_\mathrm{min}(x))$, make $\bar{x}_2$ 
		within bounds \par and compute $\bar{y}_2\gets y_\mathrm{min}(\bar{x}_2)$
		\State Set $(\bar{x},\bar{y})\gets$ best\textsuperscript{$\star$} among $(\bar{x}_1,\bar{y}_1)$ and $(\bar{x}_2,\bar{y}_2)$
	\EndIf
	\State \textbf{Output:} $(\Na,\Ma)\gets (\round{N/\bar{x}},\round{\bar{y}})$
\end{algorithmic}
\label{alg1}
\textsuperscript{$\star$}Best is the one that minimizes $f(x,y)$
\end{minipage}
\end{algorithm}

\section{Numerical Evaluation}\label{section:evaluation}

In this section, we numerically evaluate the proposed solution~(\ref{eq:sol_with_round}), which we refer to as optimized, and compare it with the 
awake-but-whisper, rush-to-sleep, and rush-to-sleep strategies. The statistical distribution of the large-scale fading coefficients 
is derived from measurements of macro sub-6 GHz 4G and 5G \glspl{bs} in Belgium~\cite{Agram_2024}, while the considered \gls{bs} configurations and 
related parameters are given in Table~\ref{tab:parameters_Pcons}.

\subsection{Large-Scale Fading and Network Load Computation}

The evaluation scenario comprises the set of large-scale fading coefficients $\left\{\beta_k\right\}$ and users' target rates $\left\{R_k\right\}$.
To compute the large-scale fading coefficients, we use the \gls{cqi} distribution obtained from 
on-site measurements of macro sub-6 GHz \glspl{bs} in Belgium~\cite{Agram_2024}. We selected the data from three sectors of two \glspl{bs}:
\begin{itemize}[leftmargin=*]
    \item A 4G \gls{lte} \gls{rru} with 4 TX/RX chains and antennas, operating at $f_\mathrm{c}=1.8$ GHz with $B=20$ MHz.
    \item A 5G \gls{nr} \gls{aau} with 8 TX/RX chains and antennas, operating at $f_\mathrm{c}=3.5$ GHz with $B=100$ MHz.
\end{itemize}
The raw data give the number of times each \gls{cqi} index (from $0$ to $15$) is reported every hour over six days, and from those we
compute the two \gls{cqi} statistical distributions (one for 4G \glspl{bs} and one for 5G \glspl{bs}). Given the use of different modulation orders, 4G and 5G systems use two different
\gls{cqi} tables~\cite[Table~5.2.2.1-2 and Table~5.2.2.1-3]{3gpp_TS_38.214}. We project all the \glspl{cqi} on the second table to perform a fair comparison.
The \gls{cqi} can then be mapped to the \gls{snr} by using the appropriate relations~\cite{Wang_2020}.
Once the \gls{snr} of user $k$ is known, denoted as $\mathrm{SNR}_k$, our rate model~(\ref{eq:rate}) shows that $\mathrm{SNR}_k=p_k/\sigma_k^2$.
Hence, the power allocation at user $k$ is given by $\sigma_k^2\mathrm{SNR}_k$. 
The noise power is computed as $\sigma_k^2=T_\sigma k_\mathrm{\scriptscriptstyle{B}} B F_\sigma$, where 
$T_\sigma=290$ K and $F_\sigma=9$ dB are the noise temperature and the noise figure, and $k_\mathrm{\scriptscriptstyle{B}}$ is the Boltzmann 
constant~\cite{ngo_cell-free_2017}.
Last, we use the expression of the total transmit power~(\ref{eq:array_gain}) to compute $\beta_k$ by approximating $K=1$, yielding
\begin{equation*}
    \beta_k = \frac{\sigma_k^2\mathrm{SNR}_k}{P_\mathrm{\scriptscriptstyle{T}}(M-1)}.
\end{equation*}
Based on the measurements and on the certificates of conformity that are publicly available upon request for deployed \glspl{bs}~\cite{BS_attest}, we set
$P_\mathrm{\scriptscriptstyle{T}}$ to $160$ W, $32$ W, and $20$ W for the 4T4T, $\mathsf{8T8R}$, and $\mathsf{64T64R}$ configurations, respectively. 

Concerning the users' target rates, let us consider some baseline rates $\left\{R_{k,0}\right\}$ normalized such that $\sum_{k=1}^K R_{k,0} = 1$,
so $\left\{R_{k,0}\right\}$ are the users' shares of the sum rate. The baseline rates are then scaled by $\kappa$ giving $R_k=\kappa R_{k,0}$. 
The maximum rate scaling $\kappa_\mathrm{max}$ is therefore the value of $\kappa$ that solves~(\ref{eq:P_max_constraint_simplified}) with equality when $\Ma=M$ and $\Na=N$
\begin{equation*}
M = \frac{K}{2}+\frac{1}{2}\sqrt{K^2+4\sum_{k=1}^K\rho_k^{-1}\left(2^{\kappa_\mathrm{max}R_{k,0}}-1\right)}
\end{equation*}
for a fixed set of $\{\beta_k\}$ and $\left\{R_{k,0}\right\}$.
With this choice and given the definition of $N_\mathrm{a,min}$ in~(\ref{eq:N_a_min}), 
$N_\mathrm{a,min} = \ceil{\left(\kappa/\kappa_\mathrm{max}\right)N}$. In the following, we generate $\{R_{k,0}\}$ as
$K$ realizations of a standard uniform distribution that are normalized to sum to $1$, and we define the network load as the ratio 
$\kappa/\kappa_\mathrm{max}$. Larger $\kappa$ indicate more loaded systems in terms of larger user rates.

\begin{figure*}[!t]
    \small
    \centering
    Configuration: $\mathsf{64T64R}$, $N=100$ time slots
    
    \vspace{.1cm}
    \resizebox{.4\linewidth}{!}{
    \includegraphics{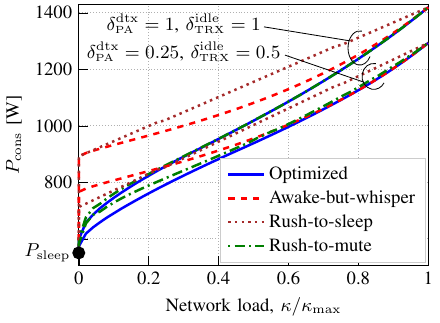}
    }
	\hfil
    \resizebox{.4\linewidth}{!}{
    \includegraphics{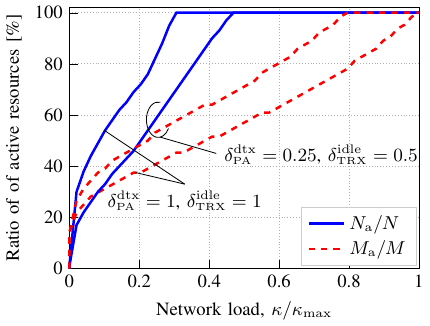}
    }
    \vspace{-.3cm}
    \caption{For $\mathsf{64T64R}$ configuration, (left) power consumption vs. network load for different energy-saving strategies, (right) optimal number of active spatial and time resources vs. network load.
    Disabled time-domain hardware power-saving modes corresponds to $\delta_\mathrm{\scriptscriptstyle{PA}}^\mathrm{dtx}=1$ and $\delta_\mathrm{\scriptscriptstyle{TRX}}^\mathrm{idle}=1$, while
    enabled time-domain hardware power-saving modes corresponds to $\delta_\mathrm{\scriptscriptstyle{PA}}^\mathrm{dtx}=0.25$ and $\delta_\mathrm{\scriptscriptstyle{TRX}}^\mathrm{idle}=0.5$.}
    \label{fig:_vs_kappa}
\end{figure*}

\begin{figure*}[!t]
    \vspace{.2cm}
    \small
    \centering
    \underline{Disabled time-domain hardware power-saving modes} ($\delta_\mathrm{\scriptscriptstyle{PA}}^\mathrm{dtx}=1$ and $\delta_\mathrm{\scriptscriptstyle{TRX}}^\mathrm{idle}=1$), $N=100$ time slots
    \vspace{.1cm}
    \includegraphics{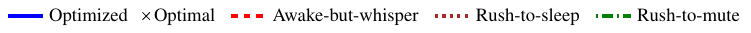}
    \resizebox{.9\linewidth}{!}{
    \includegraphics{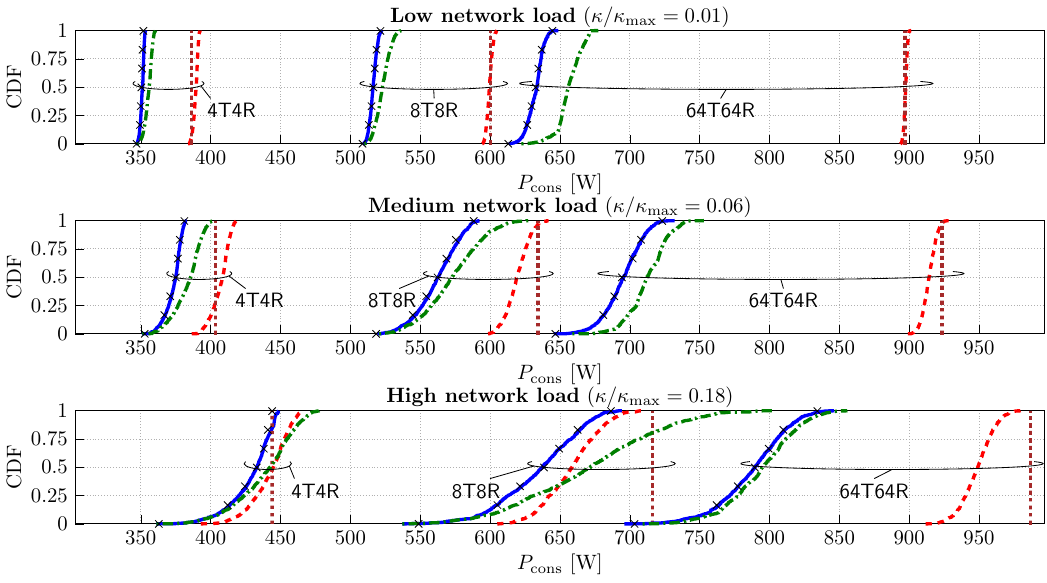}
    }
    \vspace{-.3cm}
    \caption{For disabled PA \gls{mudtx} and \gls{afe} idle-mode power savings, \glspl{cdf} of power consumption for the three \gls{bs} configurations at different network loads.}
    \label{fig:CDFs_1}
\end{figure*}

\subsection{Power Consumption and Optimal Number of Resources as a Function of Network Load}

We consider a fixed set of $\{\beta_k\}$ and $\left\{R_{k,0}\right\}$, giving a unique value of $\kappa_\mathrm{max}$, and vary $\kappa$
to observe the evolution of $P_\mathrm{cons}$ and $(\Na,\Ma)$ as the network load changes. We focus here on the massive \gls{mimo} configuration
$\mathsf{64T64R}$. We notice for this and other configurations that with the computed values of $\beta_k$, the domain $\mathcal{D}$ is often not convex. 
This and the use of the rounding operator make the proposed solution an approximation of the optimal one.
However, in many cases the solution lies on the left-edge of $\mathcal{D}$ (see Fig.~\ref{fig:set_D}), hence the non-convexity of the domain has no 
effect on the approximated solution.

The left plot of Fig.~\ref{fig:_vs_kappa} shows that at any network load, the rush-to-mute approach achieves similar or equivalent performance to
the optimized solution when there are no time-domain hardware power savings ($\delta_\mathrm{\scriptscriptstyle{PA}}^\mathrm{dtx}=1$ and $\delta_\mathrm{\scriptscriptstyle{TRX}}^\mathrm{idle}=1$). 
This is in line with the values in Table~\ref{tab:parameters_Pcons}, where $P_0$ is zero while $P_1$ assumes large values. 
All the four strategies consume $\Psleep$ at zero network load, while they consume the same amount of power at 
maximum network load as that point corresponds to the activation of all the time slots and the antennas at $\Pmax$.
The difference between the proposed or rush-to-mute scheme and the awake-but-whisper or rush-to-sleep scheme becomes smaller as the network load increases.
This is logical as most of the resources should be activated when there are large users' target rates.
When time-domain power-saving modes are enabled ($\delta_\mathrm{\scriptscriptstyle{PA}}^\mathrm{dtx}=0.25$ and $\delta_\mathrm{\scriptscriptstyle{TRX}}^\mathrm{idle}=0.5$), 
the gap between the optimized and rush-to-mute solutions increases at low network loads because the active \glspl{pa} and TR/RX chains consume less power.
Also, the awake-but-whisper becomes optimal at very high network loads.

\begin{figure*}[!t]
    \small
    \small
    \centering
    \underline{Enabled time-domain hardware power-saving modes} ($\delta_\mathrm{\scriptscriptstyle{PA}}^\mathrm{dtx}=0.25$ and $\delta_\mathrm{\scriptscriptstyle{TRX}}^\mathrm{idle}=0.5$), $N=100$ time slots
    \vspace{.1cm}
    \includegraphics{legend.pdf}
    \resizebox{.9\linewidth}{!}{
    \includegraphics{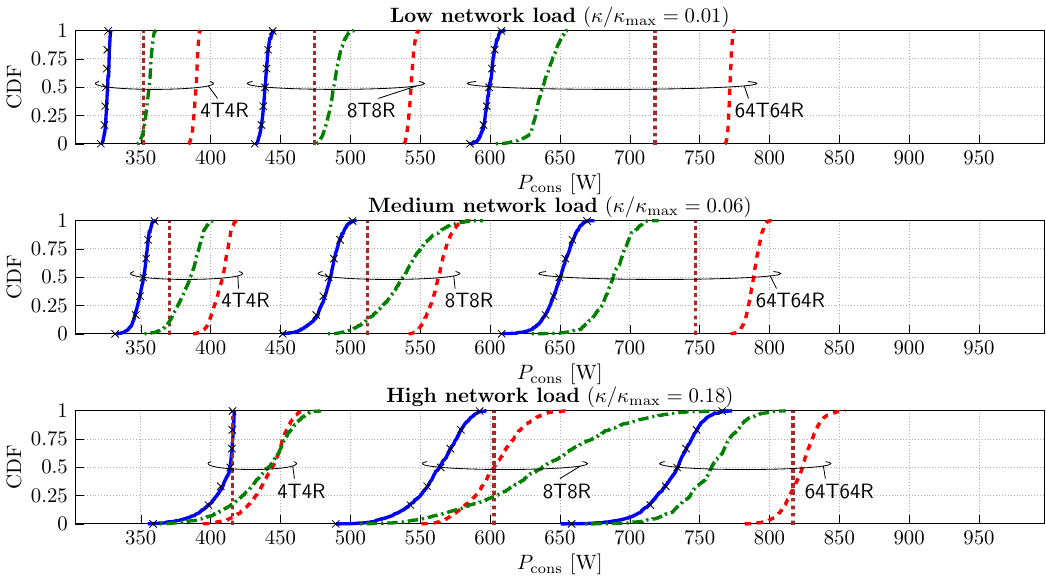}
    }
    \vspace{-.3cm}
    \caption{For enabled PA \gls{mudtx} and \gls{afe} idle-mode power savings, \glspl{cdf} of power consumption for the three \gls{bs} configurations at different network loads.}
    \label{fig:CDFs_2}
\end{figure*}

The right plot of Fig.~\ref{fig:_vs_kappa} presents the change in number of optimal resources with the network load. The value of $\Na$ first increases and then reaches its maximum value $N$ 
at a certain value of network load. On the other hand, $\Ma$ keeps increasing until the maximum network load when there are no time-domain power savings, 
while it reaches its maximum value $M$ at a traffic load below $1$ when time-domain power savings are enabled.
We observe a similar trend for different $\{\beta_k\}$ and $\left\{R_{k,0}\right\}$.
Therefore, when looking at the dependency on the network load in massive \gls{mimo} $\mathsf{64T64R}$ configuration we find that:
\begin{itemize}[leftmargin=*]
    \item When there are no time-domain hardware power-saving modes, the optimal energy-saving strategy operates first in time, space and power,
    and then in space and power only.
    \item When there are time-domain hardware power-saving modes, the optimal energy-saving strategy operates first in time, space and power,
    then in space and power only, and last in power only.
\end{itemize}

\subsection{Average Power Consumption Performance}

We now evaluate the \gls{cdf} of the power consumed by the four energy-saving schemes for the three \gls{bs} configurations at
three network load levels: low, medium, and high network load, corresponding to $\kappa/\kappa_\mathrm{max}=0.01$, 
$\kappa/\kappa_\mathrm{max}=0.06$, and $\kappa/\kappa_\mathrm{max}=0.18$, respectively.
We consider $10^3$ realizations of $\left\{\beta_k\right\}$ and $\left\{R_{k,0}\right\}$.
The \glspl{cdf} of the rush-to-sleep approach assume values only in one point given that $N_\mathrm{a,min}$ is fixed as $\ceil{\left(\kappa/\kappa_\mathrm{max}\right)N}$.
We observe that the \glspl{cdf} obtained by using the optimized solution (solid blue) and the global optimum (black crosses) present a negligible difference.

In Fig.~\ref{fig:CDFs_1} we investigate the performance when PA \textmu DTX and AFE idle-mode power savings are disabled. 
At low network load, the optimized and rush-to-mute schemes perform significantly better than the rush-to-mute and awake-but-whisper for 
any \gls{bs} configuration. The optimized solution achieves $30\%$ ($\mathsf{64T64R}$), $14\%$ ($\mathsf{8T8R}$), and $10\%$ ($\mathsf{4T4R}$) median energy savings with respect to the rush-to-sleep and awake-but-whisper. 
At medium network load, the difference between the four energy-saving schemes decreases but 
the optimized and rush-to-mute remain well separated from the other two schemes. This gap further decreases at high network load, as highlighted also previously. 
Except for the $\mathsf{64T64R}$ \gls{bs}, the rush-to-mute approach becomes more distant to the proposed scheme while the 
awake-but-whisper (for the $\mathsf{8T8R}$ \gls{bs}) and both the awake-but-whisper and rush-to-mute (for the $\mathsf{4T4R}$ \gls{bs}) reduce their gap with the proposed scheme.
We highlight that the three \gls{bs} configurations cannot be directly compared as a larger $M$ is associated with larger $\kappa_\mathrm{max}$, $K$, and sum rate.

The performance of the rush-to-sleep technique comparatively improves when PA \textmu DTX and AFE idle-mode power savings are enabled, as
shown in Fig.~\ref{fig:CDFs_2}. For the $\mathsf{64T64R}$ configuration at any network load, the rush-to-mute scheme remains the closest to the optimal one, 
with a larger difference than in Fig.~\ref{fig:CDFs_1} indicating that the optimal scheme uses a combination of time- space-, 
and power-domain energy-saving techniques. The optimized scheme achieves $17\%$, $13\%$, and $6\%$ median energy savings with respect to 
awake-but-whisper, rush-to-mute, and rush-to-sleep at medium network load. For the $\mathsf{8T8R}$ configuration, the rush-to-sleep achieves the highest performance 
among the benchmarks at low and medium network loads, while at high network load its performance is similar to the awake-but-whisper. 
The rush-to-sleep is instead the closest approach to the proposed one for the $\mathsf{4T4R}$ configuration at any network load. 
In general, even with active time-domain hardware power-saving modes, the cost of activating all the \gls{bs} antennas in massive \gls{mimo} remains high and this still favors
a rush-to-mute allocation. Different is the case for the two other configurations where the optimized solution exploits more the degrees of freedom in time.

\section{Conclusion}\label{section:conclusion}

We studied the BS resource allocation problem in time, space, and power domains under \gls{qos} and power constraints using a
state-of-the-art parametric power consumption model. The initially involved optimization problem can be simplified and solved via one, two or three unconstrained optimizations.
Practical insights are drawn for 4G and 5G \glspl{bs}. It turns out that a rush-to-mute approach, using as less antennas as possible at maximum output power, 
provides quasi-optimal performance for the analyzed \glspl{bs} that are not implementing time-domain power-saving modes. 
When the \gls{bs} hardware can enter time-domain power-saving modes, the optimal energy-saving strategy utilizes combinations of intermediate number of active time slots,
active antennas, and transmit power below the maximum one. The difference between the use of the three energy-saving domains decreases at high network loads.
Median energy savings up to $25\%$ are attained at medium network loads.
Future work includes deriving scaling regimes and optimality conditions for the three one-dimensional energy-saving allocations, 
the inclusion of frequency-domain energy-saving techniques, and the analysis of cell-free scenarios. Further, evaluating the energy savings of
the proposed scheme under actual traffic scenarios, using real hardware and considering instantaneous performance indicators over the average ones, is highly relevant.

\newcommand{\vzeta}{\vect{\zeta}}
\newcommand{\vr}{\vect{r}}
\newcommand{\vgamma}{\vect{\gamma}}
\newcommand{\vone}{\vect{1}}

\section{Appendix}
\label{section:appendix}

\subsection{Simplification of $\Pmax$ Constraint (\ref{eq:P_max_constraint})}\label{appendix:P_max_simplified}
Let us use the definition $y=\Ma$ and consider it as continuous. Using $\rho_k=\Pmax\beta_k/\sigma_k^2$, we can rewrite~(\ref{eq:P_max_constraint}) as
\begin{align*}	 
	0\leq y^2-Ky-\sum_{k=1}^{K}\rho_k^{-1} \left(2^{R_k\frac{N}{\Na}}-1\right).
\end{align*}
The right term is a quadratic expression in $y$ with two roots
\begin{align*}
	y&=\frac{K}{2}\pm\frac{1}{2}\sqrt{K^2+4\sum_{k=1}^{K}\rho_k^{-1} \left(2^{R_k\frac{N}{\Na}}-1\right)}.
\end{align*}
Given that $\Ma\geq K$, also $y\geq K$ and we should only keep the one with positive sign. 
Given that the parabolic function is looking upward, we find the condition
\begin{align*}
	y\geq \frac{K}{2}+\frac{1}{2}\sqrt{K^2+4\sum_{k=1}^{K}\rho_k^{-1} \left(2^{R_k\frac{N}{\Na}}-1\right)}
\end{align*}
which implies~(\ref{eq:P_max_constraint_simplified}) going back to the discrete domain.


\subsection{Convexity of the Function $f(x,y)$}\label{appendix:convexity_f}

We want to show that the function~(\ref{eq:f_xy}) is 
jointly convex, considering that $x\geq 1$ and $K\leq y\leq M$. Clearly, the term in $y$ is convex. 
To show the convexity of the term in $\frac{y}{x}$, we check that its Hessian is positive semidefinite.
A sufficient condition is that its determinant and trace are positive. The determinant is given by $\frac{1}{x^4}\geq 0$, 
while the trace is given by $\frac{y}{x^3}\geq 0$. 
Therefore, $\frac{y}{x}$ is convex in the domain of interest. We are left with the term $g(y)h(x)$, where
\begin{align*}
g(y)={\frac{y}{(y(y-K))^{\alpha}}},\quad h(x)={\frac{\left(\sum_{k=1}^{K}\zeta_k (e^{R_k^\prime x}-1)\right)^{\alpha}}{x}}
\end{align*}
with $R_k^\prime=R_k\ln2$ and $\zeta_k = \sigma_k^2/\beta_k$. To show convexity, let us show that the determinant and the trace of its Hessian are positive. The determinant condition requires that
\begin{align*}
	h''(x)h(x)g''(y)g(y)-(h'(x)g'(y))^2\geq0
\end{align*}
which is verified if the two following conditions hold
\begin{align*}
	g''(y)g(y) - (g'(y))^2 \geq 0, \quad
	h''(x)h(x) - (h'(x))^2 \geq 0.
\end{align*}
The trace condition is
	$h''(x)g(y)+h(x)g''(y)\geq 0$ and
given that $h(x)\geq 0$ and $g(y)\geq 0$, it is verified by checking that both functions are convex, \textit{i.e.}, $h''(x)\geq 0, g''(y)\geq 0$. Let us first study $g(y)$, with derivatives
\begin{align*}
	g'(y) 
	&= \frac{1}{y^{\alpha}(y-K)^{1+\alpha}}\left(y(1-2\alpha)-K(1-\alpha)\right)\\
	g''(y) &=\alpha\frac{2y^2(2\alpha-1)+4Ky(1-\alpha)-K^2(1-\alpha)}{y^{1+\alpha}(y-K)^{2+\alpha}}.
\end{align*}
The denominator of $g''(y)$ is positive and given that $\alpha \geq 0.5$, $2\alpha-1\geq 0$. Moreover, $y\geq K$ implies $4Ky\geq K^2$, 
hence $g''(y)\geq 0$. By developing $g''(y)g(y) - (g'(y))^2$, we obtain 
\begin{align*}
	&\frac{y^2(2\alpha-1)+y2K(1-\alpha)-K^2(1-\alpha)}{y^{2\alpha}(y-K)^{2+2\alpha}}.
\end{align*}
Again, the denominator is positive and given that $\alpha \geq 0.5$, we have $2\alpha-1\geq 0$. Moreover,  $y\geq K$ implies $y2K\geq K^2$. Hence, $g''(y)g(y) - (g'(y))^2\geq 0$. Let us now study $h(x)$ and let us define the function $k(x)=\ln h(x)$. 
Given that $h(x)=e^{k(x)}$, we have
\begin{align*}
	&h'(x)=e^{k(x)}k'(x),\ h''(x)=e^{k(x)}(k'(x))^2+e^{k(x)}k''(x)\\
	&h''(x)h(x)-(h'(x))^2=e^{2k(x)}k''(x).
\end{align*}
Therefore, it is sufficient to show that $k''(x) \geq 0$. After simplification and manipulation, $k''(x)$ can be expressed as
\begin{align*}
	k''(x) =&\; 
	\alpha \frac{\sum_{k=1}^{K}\sum_{k'=1}^{K}\zeta_k\zeta_{k'}t_{k,k'}}{\left(\sum_{k=1}^{K}\zeta_k (e^{R_k^\prime x}-1)\right)^2} + (1-\alpha) \frac{1}{x^2}.
\end{align*}
where
\begin{align*}
	t_{k,k'}&=\frac{1}{2}{R_k^\prime}^2 e^{xR_k^\prime} (e^{xR_{k'}^\prime}-1)+\frac{1}{2}(e^{xR_k^\prime}-1)e^{xR_{k'}^\prime}{R_{k'}^\prime}^2\\
	&-R_k^\prime e^{R_k^\prime x}e^{R_{k'}^\prime x}R_{k'}^\prime+\frac{1}{x^2}(e^{R_k^\prime x}-1)(e^{R_{k'}^\prime x}-1).
\end{align*}
Given that $\alpha \in [0.5,1]$, a sufficient condition for $k''(x)\geq 0$ is that $t_{k,k'}\geq 0 \ \forall k,k'$. Using $\sinh(x)\geq x$ for $x\geq 0$, we find the inequality
\begin{align*}
	(e^{xR_{k}^\prime}-1)=2e^{\frac{xR_{k}^\prime}{2}}\sinh\left({xR_{k}^\prime}/{2}\right)\geq e^{\frac{xR_{k}^\prime}{2}}{xR_{k}^\prime}
\end{align*}
and thus
\begin{align*}
	&t_{k,k'}\geq \frac{1}{2}{R_k^\prime}^2 e^{xR_k^\prime} (e^{xR_{k'}^\prime}-1)+\frac{1}{2}(e^{xR_k^\prime}-1)e^{xR_{k'}^\prime}{R_{k'}^\prime}^2 \\
	& -R_k^\prime e^{xR_k^\prime} e^{x R_{k'}^\prime}  R_{k'}^\prime+R_k^\prime e^{xR_k^\prime/2}e^{xR_{k'}^\prime/2}R_{k'}^\prime\\
	&=-\frac{1}{2}\left(R_k^\prime e^{\frac{xR_k^\prime}{2}}-e^{\frac{xR_{k'}^\prime}{2}}R_{k'}^\prime\right)^2+\frac{1}{2}e^{xR_k^\prime+xR_{k'}^\prime}\left(R_k^\prime-R_{k'}^\prime\right)^2.
\end{align*}
We need to show that this term is greater or equal than zero. Multiplying by $x^2/2$ and defining $a=xR_k/2$ and $b=xR_{k'}/2$, it is equivalent to show that
\begin{align*}
e^{2a+2b}(a-b)^2\geq \left(ae^{a}-be^{b}\right)^2.
\end{align*}
Let us consider that $a>b$ (otherwise symmetric) and take the square root, we should then show that

$$e^{a+b}(a-b)\geq ae^{a}-be^{b}.$$ 
Subtracting $(a-b)e^a$ gives $(e^{a+b}-e^a)(a-b)\geq b(e^{a}-e^b)$,
and dividing by $e^a$, we obtain 
$(e^{b}-1)(a-b)\geq b(1-e^{b-a})$.
We can then exploit $b \leq e^b-1$ and $1-e^{b-a}\leq a-b$ to arrive at the result.

\subsection{Convexity of the Domain $\mathcal{D}$}\label{appendix:convexity_D}
The linear constraints in the definition of $\mathcal{D}$ in~(\ref{eq:set_D}) are clearly convex. Let us look further at the $\Pmax$ inequality, 
which can be rewritten as $-y+\frac{K}{2}+m(x)\leq 0$, where
\begin{align*}
	m(x)=\sqrt{\frac{K^2}{4}+\sum_{k=1}^{K}\rho_k^{-1}(e^{R_k'x}-1) }.
\end{align*}
Given that $-y$ is convex, we need to check when $m''(x)\geq 0$. Let us further define
\begin{align*}
	l(x)&=\sum_{k=1}^{K}\rho_k^{-1}\left(e^{R_k'x}+\frac{K}{4}\rho_k-1\right)
\end{align*}
so that $m(x)=\sqrt{l(x)}$ with derivatives
\begin{align*}
	m'(x)=\frac{1}{2}\frac{l'(x)}{\sqrt{l(x)}},\ m''(x)
	=\frac{1}{4}\frac{2l''(x)l(x)-(l'(x))^2}{(l(x))^{3/2}}.
\end{align*}
Given that $l(x)\geq 0$, we need to check when $2l''(x)l(x)-(l'(x))^2\geq 0$. We have
\begin{align*}
	&2l''(x)l(x)-(l'(x))^2=\\
	&\sum_{k=1}^K\sum_{k'=1}^K\frac{R_k2^{R_kx}}{\rho_k\rho_{k'}}\left[2R_k\left(\frac{K}{4}\rho_{k'}-1\right)+2^{R_{k'}x}(2R_k-R_{k'})\right].
\end{align*}
The double sum contains $K^2$ additive terms which can be partitioned in $K$ terms where $k=k'$ and $K(K-1)$ terms where $k\neq k'$. In the following we show that the condition~(\ref{eq:convexity_constraint}) which can be rewritten as
\begin{align}\label{eq:condition_rho_k}
	\rho_k\geq \frac{2}{K} \iff \frac{K}{4}\rho_{k}-1\geq -\frac{1}{2}
\end{align}
for $k=1,\dotsc,K$ is sufficient to guarantee that $k''(x)\geq 0$. Let us first consider the $K$ terms of the double sum where $k=k'$. These are positive if
\begin{align*}
		2R_k\left(\frac{K}{4}\rho_{k}-1\right)+2^{R_{k}x}(2R_k-R_{k})&\geq 0\\
		-2\left(\frac{K}{4}\rho_{k}-1\right)&\leq 2^{R_{k}x}
\end{align*}
which is verified if~(\ref{eq:condition_rho_k}) holds. Let us now consider the $K(K-1)$ terms where $k\neq k'$. For each term $k,k'$, there exists also a term $k',k$. 
The total set of $K(K-1)$ terms can then be subdivided in $K(K-1)/2$ pairs of distinct indices $k,k'$. For instance, if $K=3$, there exists $3(3-1)/2=3$ pairs: 
$1,2$ and $2,1$; $1,3$ and $3,1$; $2,3$ and $3,2$. Let us consider the addition of one pair of terms $k,k'$ and $k',k$. The contribution of these two terms is 
positive if $f_{k,k'}\geq 0$ where

\begin{align*}
	f_{k,k'}&=R_k2^{R_kx}\left[2R_k\left(\frac{K}{4}\rho_{k'}-1\right)+2^{R_{k'}x}(2R_k-R_{k'})\right]\\
	&+R_{k'}2^{R_{k'}x}\left[2R_{k'}\left(\frac{K}{4}\rho_{k}-1\right)+2^{R_{k}x}(2R_{k'}-R_{k})\right].
\end{align*}
If~(\ref{eq:condition_rho_k}) holds, we have
\begin{align*}
	&f_{k,k'}\geq R_k2^{R_kx}\left[-R_k+2^{R_{k'}x}(2R_k-R_{k'})\right]\\
	&\quad \ \ \ +R_{k'}2^{R_{k'}x}\left[-R_{k'}+2^{R_{k}x}(2R_{k'}-R_{k})\right]\\
	&=2^{R_kx+R_{k'}x}\left[2R_k^2+2R_{k'}^2-2R_{k'}R_k-\frac{R_k^2}{2^{R_{k'}x}}-\frac{R_{k'}^2}{2^{R_{k}x}}\right]
\end{align*}
and given that $-1/2^{R_{k}x} \geq -1$ and $-1/2^{R_{k'}x} \geq -1$, we find
\begin{align*}
	f_{k,k'}&\geq 2^{R_kx+R_{k'}x}\left[2R_k^2+2R_{k'}^2-2R_{k'}R_k-R_k^2-R_{k'}^2\right]\\
	&= 2^{R_kx+R_{k'}x}(R_k-R_{k'})^2
\end{align*}
which is positive. Hence, the contribution from all pairs of indices where $k\neq k'$ gives a positive contribution to the double sum. The same holds for the terms where $k=k'$. The proof is thus complete.

\subsection{Error Order of Using a Rounding Operation}\label{appendix:ceil_floor}

Let us define the optimal allocation as $(\Na,\Ma)=\big\lfloor N/\bar{x},\bar{y}\big\rceil$ and its approximation through the rounding operator as 
$\hat{N}_{\mathrm{a}}=[N/\bar{x}],\ \hat{M}_{\mathrm{a}}= [\bar{y}]$. We can also write
\begin{align*}
	\hat{N}_{\mathrm{a}}=\Na+\epsilon_1,\ \hat{M}_{\mathrm{a}}=\Ma+\epsilon_2
\end{align*}
where $|\epsilon_1|\leq 1$, $|\epsilon_2|\leq 1$. The extra consumed power of the approximation with respect to the optimal allocation is defined
\begin{align*}
	\epsilon_3={P}_{\mathrm{cons}}(\hat{N}_{\mathrm{a}},\hat{M}_{\mathrm{a}})-\Pcons(\Na,\Ma) \label{eq:def_epsilon_3}
\end{align*}
where the function $\Pcons(\Na,\Ma)$ is given in~(\ref{eq:simplified_prob2}). 
Developing $\epsilon_3$, we find
\begin{align*}
	\epsilon_3
	&=\frac{\epsilon_2}{M}P_1+\left(\Na\epsilon_2+\epsilon_1\Ma+\epsilon_1\epsilon_2\right)\frac{P_0}{NM}  \\ 
	&+\frac{\Na+\epsilon_1}{N}(\Ma+\epsilon_2)\gamma \left(\frac{\sum_{k=1}^{K}\frac{\sigma_k^2}{\beta_k} \left(2^{R_k\frac{N}{\Na+\epsilon_1}}-1\right)}{(\Ma+\epsilon_2)(\Ma+\epsilon_2-K)}\right)^{\alpha}\\
	& - \frac{\Na}{N}\Ma\gamma \left(\frac{\sum_{k=1}^{K}\frac{\sigma_k^2}{\beta_k} \left(2^{R_k\frac{N}{\Na}}-1\right)}{\Ma(\Ma-K)}\right)^{\alpha}.
\end{align*}
As $\Ma \rightarrow +\infty$ and $N\rightarrow +\infty$, the first two terms can be bounded as $\mathcal{O}(1/N+1/M)$. For the two last ones, after performing a first order Taylor expansion around $\epsilon_1=0$ and $\epsilon_2=0$, 
we find the bound $\mathcal{O}(1/N+1/\Ma^{2\alpha})$ so that $|\epsilon_3|=\mathcal{O}(1/N+1/M+1/\Ma^{2\alpha})$.
\section*{Acknowledgment}
The authors would like to thank Louis Golard for elucidating the details of the power consumption model.

\ifCLASSOPTIONcaptionsoff
  \newpage
\fi



\bibliographystyle{IEEEtran}
\bibliography{IEEEabrv,refs}

\end{document}